\documentclass[prd,aps,showpacs,tightenlines,10pt,reprint,nofootinbib,superscriptaddress, floatfix,twocolumn]{revtex4-2}

\usepackage[utf8]{inputenc} 

\DeclareUnicodeCharacter{2212}{-}
\usepackage{acronym}
\usepackage{amsmath}
\usepackage{amsthm}  
\usepackage{amssymb}
\usepackage{mathtools}
\usepackage{phaistos}
\usepackage{graphicx}
\usepackage[font=small,labelfont=bf,justification=Justified]{caption}
\usepackage{subcaption}
\usepackage[section]{placeins}

\usepackage{physics} 
\usepackage{IEEEtrantools}  
\usepackage{xspace}
\usepackage{cancel}
\usepackage[normalem]{ulem} 
\usepackage{comment}
\usepackage{epsfig}
\usepackage{multirow} 
\usepackage{color}
\usepackage[dvipsnames]{xcolor}
\usepackage{rotating}
\usepackage[export]{adjustbox} 
\usepackage{floatrow}
\newfloatcommand{capbtabbox}{table}[][\FBwidth]
\usepackage{tabularx}
\usepackage[pdfencoding=auto, psdextra]{hyperref}
\usepackage{units}
\usepackage[T1]{fontenc}
\usepackage[normalem]{ulem} 
\usepackage{afterpage}

\usepackage{tcolorbox} 

\hypersetup{
    colorlinks,
    linkcolor={red!50!black},
    citecolor={blue!50!black},
    urlcolor={blue!80!black}
}

\newcommand{\RD}[1]{\textcolor{teal}{#1}}

\newcommand{\figref}[1]{Fig.~\ref{#1}}
\renewcommand{\eqref}[1]{Eq.~(\ref{#1})}

\begin{document}

\title{Tidal Deformability of Fermion-Boson Stars:\texorpdfstring{\\}{} Neutron Stars Admixed with Ultra-Light Dark Matter}

\author{Robin Fynn Diedrichs}
\email{diedrichs@itp.uni-frankfurt.de}
\affiliation{Institute for Theoretical Physics, Goethe University, 60438 Frankfurt am Main, Germany}

\author{Niklas Becker}
\email{nbecker@itp.uni-frankfurt.de}
\affiliation{Institute for Theoretical Physics, Goethe University, 60438 Frankfurt am Main, Germany}

\author{Cédric Jockel}
\email{jockel@itp.uni-frankfurt.de}
\affiliation{Institute for Theoretical Physics, Goethe University, 60438 Frankfurt am Main, Germany}

\author{Jan-Erik Christian}
\email{christian@itp.uni-frankfurt.de}
\affiliation{Institute for Theoretical Physics, Goethe University, 60438 Frankfurt am Main, Germany}

\author{Laura Sagunski}
\email{sagunski@itp.uni-frankfurt.de}
\affiliation{Institute for Theoretical Physics, Goethe University, 60438 Frankfurt am Main, Germany}

\author{Jürgen Schaffner-Bielich}
\email{schaffne@itp.uni-frankfurt.de}
\affiliation{Institute for Theoretical Physics, Goethe University, 60438 Frankfurt am Main, Germany}

\date{\today}

\begin{abstract}
In this work we investigate the tidal deformability of a neutron star admixed with dark matter, modeled as a massive, self-interacting, complex scalar field. We derive the equations to compute the tidal deformability of the full Einstein-Hilbert-Klein-Gordon system self-consistently, and probe the influence of the scalar field mass and self-interaction strength on the total mass and tidal properties of the combined system. We find that dark matter core-like configurations lead to more compact objects with smaller tidal deformability, and dark matter cloud-like configurations lead to larger tidal deformability. Electromagnetic observations of certain cloud-like configurations would appear to violate the Buchdahl limit. The self-interaction strength is found to have a significant effect on both mass and tidal deformability. We discuss observational constraints and the connection to anomalous detections. We also investigate how this model compares to those with an effective bosonic equation of state and find the interaction strength where they converge sufficiently. 

\end{abstract}

\maketitle

\section{Introduction \label{sec:intro}} 
Neutron stars are highly compact remnants of massive stars. Due to the high densities inside of neutron stars, they allow us to probe nuclear matter at high densities, a region that is not readily accessible with analytic techniques. 

The equation of state (EoS) describes the interplay between density and pressure, which is needed to close the Tolman-Oppenheimer-Volkoff (TOV) equations~\cite{Tolman:1939:PhysRev.55.364, Oppenheimer:1939:PhysRev.55.374} that describe the density profile of a spherically symmetric star and the curvature of space-time that is produced self-consistently. A significant constraint on the EoS is the mass value of the most massive known compact star. If an EoS is not able to generate a star of this mass, it cannot describe reality. There are multiple pulsars with masses at or above $2$\,M$_\odot$~\cite{Demorest:2010bx,Antoniadis:2013pzd,Fonseca:2016tux,Cromartie:2019kug,Nieder:2020yqy}. Recently even a $2.35\pm0.17$\,M$_\odot$ neutron star was reported by Romani et al.~\cite{Romani:2022jhd}. There is also some speculation that the lighter companion of the GW190814 gravitational wave event \cite{Abbott_2020} was the most massive neutron star ever observed, with a mass of about 2.6\,M$_\odot$. However, there is some evidence that the object should be considered the lightest observed black hole instead \cite{Most:2020bba,Fattoyev:2020cws,Dexheimer:2020rlp,Tews:2020ylw,Blaschke:2020vuy,Nathanail:2021tay}. Such high masses require stiff EoSs, where the energy density strongly rises with increasing pressure. This constraint is supported by the NICER measurements of the pulsars J0030+0451 \cite{Miller:2019cac,Riley:2019yda,Raaijmakers:2019qny} and J0740+6620 \cite{Miller:2021qha,Riley:2021pdl,Raaijmakers:2021uju}, which report quite large radii. The contrary is true for the neutron star merger event GW170817 detected by LIGO/Virgo \cite{TheLIGOScientific:2017qsa,Abbott:2018exr,Abbott:2018wiz}, which favors more compact configurations generated by soft EoSs.

It is additionally possible that neutron stars accumulate dark matter (DM) in a sufficient abundance to modify their observables, such as the mass, radius, and tidal deformability. These quantities have been measured in recent observations made by, e.g.,~NICER \cite{Miller:2019cac,Riley:2019yda,Raaijmakers:2019qny,Miller:2021qha,Riley:2021pdl,Raaijmakers:2021uju} and the LIGO/Virgo/Kagra(LVK) collaborations, which thus allow to constraint the properties of DM.
DM is an integral part of the $\Lambda$CDM model, which is the concordant model of cosmology \cite{Planck:2018vyg}. Despite decades of searches, its nature and properties are still largely unknown \cite{Bertone:2004pz, ParticleDataGroup:2022pth}. 
A possible contender is DM being made up of an additional scalar field in the universe \cite{Khlopov:1985jw, Ferreira:2020fam}.
The connection between neutron stars -- where the highest densities of matter are expected -- and DM has also been explored in numerous publications and is an active area of research
\cite{Goldman:1989nd,Kouvaris:2007ay,Kouvaris:2010vv,Sandin:2008db,Ciarcelluti:2010ji,Leung:2011zz,Guver:2012ba,Li:2012ii,Xiang:2013xwa,Tolos:2015qra,Mukhopadhyay:2016dsg,Ellis:2018bkr,McKeen:2018xwc,Baym:2018ljz,Motta:2018rxp,Motta:2018bil,Ivanytskyi:2019wxd,Bell:2020obw,Husain:2022bxl,Berryman:2022zic,Cassing:2022tnn}. 
DM as a scalar field could be around neutron stars as a \textit{cloud} or inside neutron stars as a \textit{core}. Neutron stars with DM cores could form 1) from a DM `seed' through accretion of baryonic matter~\cite{Ellis:2017jgp}, 2) through mergers of neutron stars and boson stars, 3) through accretion and subsequent accumulation of DM inside the neutron star~\cite{Goldman:1989nd,Kouvaris:2007ay,Kouvaris:2010vv,Ciarcelluti:2010ji,Guver:2012ba, Ivanytskyi:2019wxd,Bell:2020obw} or 4) through the decay of standard model particles inside the neutron star into DM~\cite{Baym:2018ljz,Motta:2018rxp, Motta:2018bil,Husain:2022bxl,Berryman:2022zic}.
The presence of DM clouds and cores in and around neutron stars will affect the observable properties of the neutron stars, thus making them indirect laboratories for DM properties. It was previously shown that even large dark matter fractions of up to 20\% are not excluded from current observations \cite{Rutherford:2022xeb}. Present and future gravitational wave detectors have the potential to detect the possible presence of DM in merging neutron stars and to constrain the properties of DM, such as its mass and its self-interaction strength
\cite{Ellis:2017jgp,Nelson:2018xtr,Horowitz:2019aim,Bauswein:2020kor,Dengler:2021qcq,Karkevandi:2021ygv,Cardoso:2016oxy,Maselli:2017vfi,Maselli:2017vfi,Maselli:2017tfq,Mark:2017dnq,Gresham:2018rqo,Toubiana:2020lzd,Wystub:2021qrn,Emma:2022xjs,Hippert:2022snq, Goldman:1989nd, Ellis:2018bkr, Das:2018frc, Kain:2021hpk}.

In this work, we model DM as a minimally coupled complex scalar field that only interacts with the standard model (SM) via gravity and study its impact on the neutron star observables. To this end, we construct equilibrium solutions and their first-order perturbations and solve the coupled Einstein-Hilbert-Klein-Gordon (EHKG) system of equations. Such systems, termed fermion-boson stars (FBS), were first introduced by Henriques et al.~\cite{HENRIQUES198999} and subsequently analytically studied in terms of stability under radial perturbations~\cite{HENRIQUES1990511}. In \cite{DiGiovanni:2021ejn} they were connected to current constraints on the mass and radii of NSs and their dynamical properties were explored in \cite{Valdez_Alvarado_2013, Valdez_Alvarado_2020, Di_Giovanni_2020, Di_Giovanni_2021}. In all of these cases, these systems were investigated using a perfect fluid for the nuclear matter and a classical scalar field for the bosonic DM, which is an approach that we will also follow in this work. The described system is closely related to boson stars \cite{Kaup:1968zz, Ruffini:1969qy, Colpi:1986ye}, as it can be seen as a boson star that coexists with a neutron star at the same location in space.

The tidal deformability of such systems was first investigated in \cite{Nelson:2018xtr}, where the authors considered scalar bosonic DM with masses in the MeV to GeV range, which is gauged by a U(1) vector boson and focused on the parameter space that results in the formation of a dark halo. They further constructed an EoS for the bosonic sector by using mean field theory. In order to obtain solutions for their system, they thus extended the TOV equations to account for two fluids at the same time. This model was subsequently further investigated first in \cite{Rutherford:2022xeb} in terms of detectability prospects and in \cite{Giangrandi:2022wht}, where the resulting tidal deformability was presented for a wider range of parameters that also include scenarios in which the DM form a core. Similarly, in \cite{Karkevandi:2021ygv, Leung:2022wcf}, scalar DM that self-interacts via a quartic coupling was considered. Here, the authors used an effective EoS that was first derived in \cite{Colpi:1986ye} and then also used the two-fluid approach. The utilized EoS is however only valid if the self-interactions are sufficiently strong. 

Our method is also applicable to scalar fields with weak to no self-interactions. First, we review the equilibrium solutions and show the effects of the ultralight scalar field on the mass-radius relations. Then, we derive the relevant equations for the tidal deformability, and show the results of our numerical investigation. We find two classes of solutions, DM \textit{cores} inside the neutron star and DM \textit{clouds} enveloping the neutron star matter. Core solutions have higher compactness and lower tidal deformabilities, while cloud solutions can have large tidal deformabilities and lower the compactness of the overall object. For large DM fractions, a neutron star inside these clouds could appear to violate the Buchdahl limit, if the bosonic component is not observed, as for example in NICER observations. The large tidal deformabilities of the cloud solutions would be observable in LVK observations even for small DM fractions, while DM core solutions only have small effects on the tidal deformability, which would be difficult to discern from EoS effects.

This paper is structured as follows: In section \ref{sec:equilibrium_solutions} we present the construction of equilibrium solutions and further extend these equations in section \ref{sec:tidal_deformability} to also include the first-order perturbations. In section \ref{sec:results} we present the resulting tidal deformability and compare it to observational constraints. In section \ref{sec:comparison} we compare the EHKG solutions to the two-fluid model. Finally, in section \ref{sec:conclusions} we summarize our findings. Throughout this work, we use units in which $G = {\rm M}_\odot = c = 1$. See also appendix \ref{app::units} for information on the unit conversion.

\section{Equilibrium Solutions \label{sec:equilibrium_solutions}}

\begin{figure*}
    \includegraphics[width=0.49\textwidth]{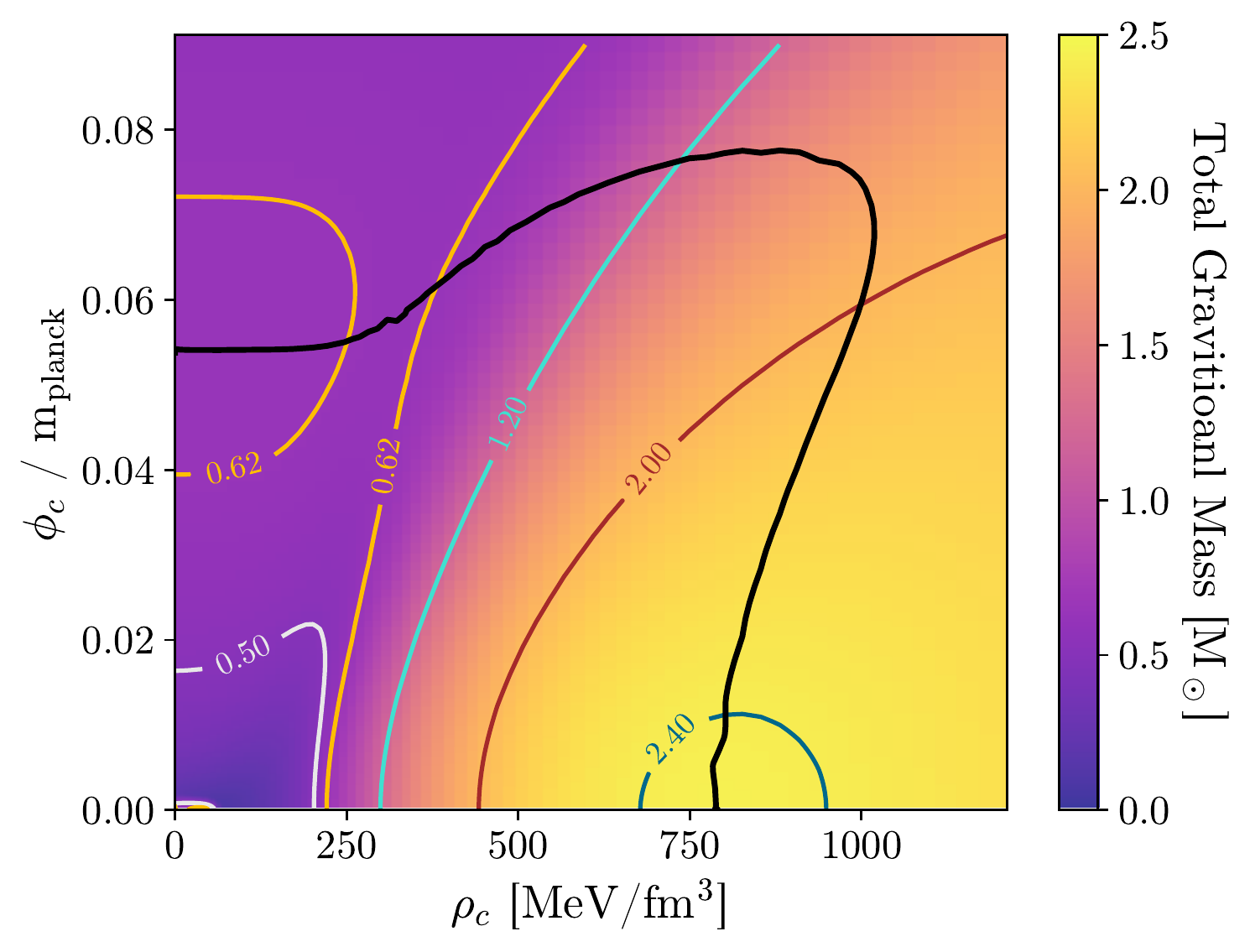}
    \includegraphics[width=0.49\textwidth]{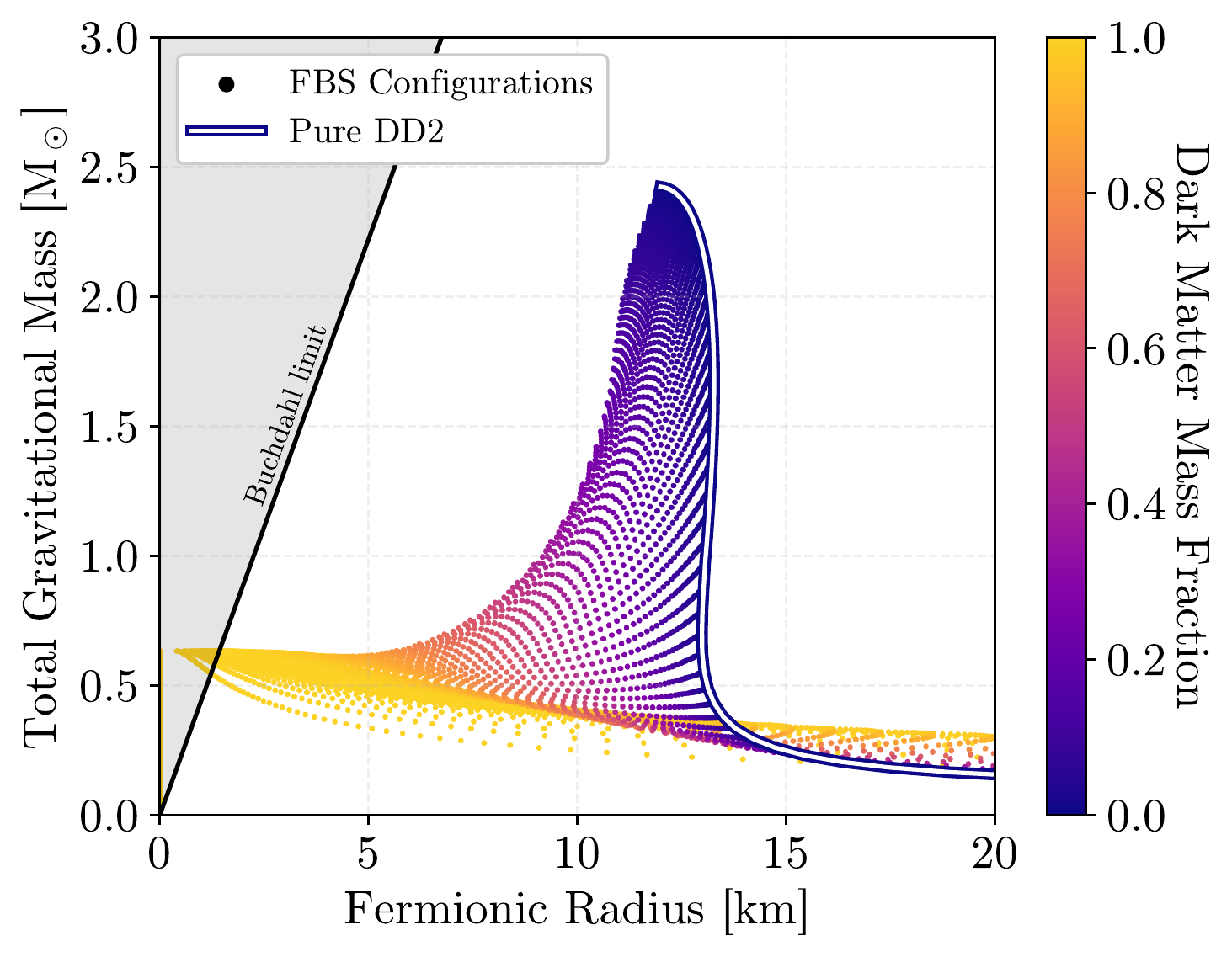}
    \caption{\textbf{Left panel:} Density plot that displays the total gravitational mass as a function of the central rest-mass density ($\rho_c$) and central value of the scalar field ($\phi_c$). Additionally, it displays the stability curve as the solid black line calculated using \eqref{stabilityCriterion}, i.e.~all configurations that lie within the bottom left parameter region that is bordered by the black line are stable against radial perturbations. 
    \textbf{Right panel:} Mass-radius diagram displaying the fermionic radius (the radius of the fermionic component) vs the total gravitational mass for configurations that are within the stable region displayed in the left panel. Each point corresponds to a single configuration and is color-coded according to the rest mass fraction of the dark matter component. The solid black line shows the mass-radius curve for pure fermionic matter. For both plots a massive scalar field with no self-interactions and the mass set to $m = 1.3 \times 10^{-10}$\,eV was considered in addition to the DD2 EoS.}
    \label{equilibrium::stabilityAndMR}
\end{figure*}

In this section, we review the construction of equilibrium solutions of FBS, which was first presented in \cite{HENRIQUES198999}. We model dark matter as a massive and complex scalar field that only interacts with the SM via gravity, such that the action of the combined system is given by

\begin{align}
    S = \int d^4x \sqrt{-g} \left[ \frac{R}{16 \pi} - \nabla_\alpha \bar{\Phi} \nabla^\alpha \Phi - V(\bar{\Phi} \Phi) + \mathcal{L}_m \right],
\raisetag{2.2\normalbaselineskip}
\end{align}

\noindent
where $\mathcal{L}_m$ is the Lagrangian describing nuclear matter and $V(\bar{\Phi} \Phi)$ is the scalar field's potential. The scalar field is invariant under a global U(1) symmetry that gives rise to a conserved Noether current

\begin{equation}
    j_\mu = i \left( \bar{\Phi} \nabla_\mu \Phi - \Phi \nabla_\mu \bar{\Phi} \right),
\end{equation}

\noindent
which allows to generally define the total number of bosons in the system as

\begin{equation}
    N_\text{b} \equiv \int d^3x \sqrt{-g} g^{0\mu} j_\mu.
\end{equation}

The energy-momentum tensor for the scalar part is given by

\begin{align}
\label{scalarfield::EMT}
    \begin{split}
        T_{\mu \nu}^{(\Phi)} = - g_{\mu \nu} \left( \partial_\alpha \bar{\Phi} \partial^\alpha \Phi + V(\bar{\Phi} \Phi) \right) \\ 
        + \partial_\mu \bar{\Phi} \partial_\nu \Phi + \partial_\mu \Phi \partial_\nu \bar{\Phi}.
    \end{split}
\end{align}

Varying the action with respect to the scalar field results in the Klein-Gordon equation, 

\begin{equation}
\label{KleinGordonEquation}
    \nabla_\mu \nabla^\mu \Phi = \Phi V'(\bar{\Phi}\Phi), \; \text{with} \; V'(\bar{\Phi}\Phi) := \frac{d V}{d |\Phi|^2}.
\end{equation}

\noindent
This equation directly implies that the energy-momentum tensor of the scalar field is separately conserved from the perfect fluid energy-momentum tensor.

\noindent
The energy-momentum tensor for nuclear matter is assumed to be of the perfect fluid form:

\begin{equation}
    T_{\mu\nu}^{\text{(NS)}} = [\rho ( 1 + \epsilon) + P] u_\mu u_\nu + P g_{\mu \nu},
\end{equation}

\noindent
where $\rho$ is the rest-mass energy density and $\epsilon$ is the internal energy density, such that $\rho(1 + \epsilon)$ describes the total energy density $e$. Requiring that the Noether current is conserved, i.e.~$\nabla_\mu (\rho u^\mu) = 0$, allows to define the total number of baryons in the system generally as

\begin{equation}
    N_\text{f} = \int d^3 x \sqrt{-g} g^{0\mu} \rho u_\mu.
\end{equation}

We consider the system (for now) to be in spherical symmetric equilibrium, such that the metric can be written as

\begin{equation}
    g_{\mu \nu} = \text{diag} \left( - e^{v(r)}, e^{u(r)}, r^2, r^2 \sin^2\theta \right).
\end{equation}

We further consider a static perfect fluid, such that $u_\mu = (e^{v/2}, 0, 0, 0)$ and write the scalar field as

\begin{equation}
    \Phi(t, r) = \phi_0(r) e^{- i \omega t},
\end{equation}

\noindent
Using the spherical symmetric ansatz together with the Klein-Gordon equation results in an equation describing the radial dependence of the bosonic field

\begin{equation}
    \phi_0'' = e^u\left(  V'(\phi_0^2) - \omega^2 e^{- v} \right) \phi_0 + \left( \frac{u' - v'}{2} - \frac{2}{r} \right) \phi_0'. \label{eq:klein-gordon-eqation}
\end{equation}

Additionally, the Einstein equations simplify to the following two equations regarding the metric functions $u(r)$ and $v(r)$:

\begin{align}
    \begin{split}
        u' &= 8 \pi r e^u \Big[ \omega^2 \phi_0^2 e^{- v} + V(\phi \bar{\phi}) \\ 
        &\hspace{2.5cm}+ e^{-u} \phi_0'^2 + \rho (1 + \epsilon) \Big] - \frac{e^u - 1}{r},
    \end{split} \\
    \begin{split}
        v' &= 8 \pi r e^u \Big[ \omega^2 \phi_0^2 e^{-v} - V(\phi \bar{\phi}) \\
        &\hspace{2.5cm}+ e^{-u} \phi_0'^2 + P \Big] + \frac{e^u - 1}{r}.
    \end{split}
\end{align}

\noindent
Also, the conservation of the energy-momentum tensor of nuclear matter $\nabla_\mu T^{\mu\nu \, \text{(NS)}} = 0$ provides a differential equation for $P$:

\begin{equation}
    P' = - [\rho (1 + \epsilon) + P] \frac{v'}{2} \label{eq:dPdr-equation}
\end{equation}

\noindent
This system of equations is closed by providing an EoS $P(\rho,\epsilon)$ (or $P(e)$) for the nuclear matter part.

Further, for the considered system, the expressions for the total number of fermions (nuclear matter) and bosons (dark matter) simplify to

\begin{align}
    N_\text{b} &= 8 \pi \int_0^\infty dr r^2 e^{(u - v) / 2} \omega \phi_0^2, \\
    N_\text{f} &= 4 \pi \int_0^{R_\text{f}} dr r^2 e^{u/2} \rho,
\end{align}

\noindent
where $R_\text{f}$ denotes the fermionic radius, which is determined by the radial position at which the fermionic pressure $P$ vanishes. The total gravitational mass of the system is given by 

\begin{equation}
    M_\text{tot} = \lim_{r \rightarrow \infty} \frac{r}{2} \left( 1 - e^{-u(r)} \right).
\end{equation}

In order to integrate these equations, it is still necessary to provide suitable initial conditions. We do this by enforcing asymptotic flatness and regularity at the origin, i.e.

\begin{align}
    \begin{split}
        \lim_{r \rightarrow \infty} v(r) &= 0, \quad\quad v(0) = v_c, \\
        \lim_{r \rightarrow \infty} u(r) &= 0, \quad\quad u(0) = 0, \\
        \lim_{r \rightarrow \infty} \phi_0(r) &= 0, \quad\quad\hskip-1.7mm \phi_0(0) = \phi_c,\\
        \phi_0'(0) &= 0, \quad\quad \rho(0) = \rho_c.
    \end{split} \label{eq:initial-conditions-full-system}
\end{align}

\noindent
Asymptotic flatness generally requires fine-tuning $v_c$ to some non-zero value. However, as was also discussed in \cite{HENRIQUES198999}, it is possible to absorb a constant shift in $v(r)$ (e.g.~$v \rightarrow v' = v - v_c$) by rescaling the above set of equations by $\omega \rightarrow \omega' = \omega e^{-v_c / 2}$. This rescaling leaves the set of equations invariant and has the advantage that we automatically have $v_c = 0$. After integrating to obtain a solution, we can retrieve the physical values of $\omega$ and $v$ by doing the inverse transformation using the asymptotic value of $v(r)$.

For given $\rho_c, \phi_c$ it is necessary to find the value of $\omega$, such that the boundary conditions at infinity (\eqref{eq:initial-conditions-full-system}) are fulfilled, i.e.\ the eigenvalues. There are infinitely many eigenvalues, which are characterized by how many nodes (i.e.~radial positions with $\phi(r) = 0$) are present in the scalar field profile. We find the lowest eigenvalue, such that there are no nodes. The bosonic ODE system is such that it will always diverge at finite radii, due to finite numerical precision. We employ this in order to efficiently find solutions. We use the fact that the scalar field profile either diverges towards $+\infty$ or $-\infty$ and changes its direction of divergence when $\omega$ passes an eigenvalue. This provides us with a binary criterion and thus allows us to implement a bisection for $\omega$ which converges exponentially fast. 

Once a sufficiently accurate $\omega$ is found, we modify the integration, such that $\phi_0$ is set to zero at a finite radius $r_{B}^*$. This radius $r_{B}^*$ is found by the condition $\phi_0 (r_{B}^*)/ \phi_c < 10^{-4}$. This is necessary because otherwise, the numerical integration diverges at finite radii. Since we have the additional neutron matter component, in some part of the parameter space, the integration would diverge before $P$ has converged to $0$. For example, in compact dark matter core configurations, the integration could diverge while still inside the neutron star component. Therefore, we artificially set $\phi_0 = 0$ for $r > r_{B}^*$, which allows us to circumvent the divergence and accurately resolve the rest of the neutron star component. The condition was chosen such that the remaining contribution of the scalar field to the other quantities (i.e. the metric components) would be minimized. We have checked for lower thresholds and the extracted results are the same. 

We integrate the system until a radius is reached at which both the scalar field $\phi_0$ and the fermionic component parameterized by $P$ have converged to zero. There, we can extract the properties of interest, such as the total mass $M_\text{tot}$ and number densities $N_f, N_b$.

For some configurations, due to numerical precision limits, the scalar field convergence condition cannot be fulfilled. This generally happens for small initial field values $\phi_c \lesssim 10^{-4}$, where the bosonic cloud extends far outside the neutron star. In these cases, we extract the mass $M_\text{tot}$ at the point where its derivative has a global minimum. When the scalar field diverges, also the metric components do, and with it the calculated mass of the system. By taking the point where the derivative of the mass has a global minimum, which roughly corresponds to where the scalar field and its derivative is closest to zero, we get the best estimate of the mass of the system before the divergence. 

Once we have a Fermion-Boson-Star solution for given $\rho_c, \phi_c$, the stability of the physical system is of importance. To this end, we need to calculate a whole family of solutions and use the stability criterion derived in \cite{HENRIQUES1990511}, which is a generalization of the stability criterion for neutron stars. The idea is to find extrema in the total number of particles for fixed mass, depending on the central values $\rho_c, \phi_c$. At these lines must be the transition between stable and unstable configurations 
\begin{equation}
\label{stabilityCriterion}
    \frac{d N_\text{f}}{d \sigma} = \frac{d N_\text{b}}{d \sigma} = 0, 
\end{equation}

\noindent
where $d/d\sigma$ denotes the derivative in the direction of constant total mass, i.e.~up to a normalization factor

\begin{equation}
    \frac{d N_\text{f}}{d\sigma} \propto - \frac{\partial M_\text{tot}}{\partial \rho_c} \frac{\partial N_\text{f}}{\partial \phi_c} + \frac{\partial M_\text{tot}}{\partial \phi_c} \frac{\partial N_\text{f}}{\partial \rho_c}.
\end{equation}

Figure \ref{equilibrium::stabilityAndMR} shows what configurations are stable depending on the central value of the rest mass density and the central value of the scalar field according to the above condition for the case of a massive scalar field with no self-interactions and the mass set to $m = 1.34 \times 10^{-10}$\,eV. Additionally, the resulting mass and radii for the stable configurations are also displayed.

\section{Tidal Deformability \label{sec:tidal_deformability}}

In order to obtain the tidal deformability, we will follow the same procedure that was used in \cite{Hinderer:2007mb} to obtain the tidal deformability of pure neutron stars and subsequently also applied to pure boson stars in \cite{Mendes:2016vdr, Sennett:2017etc}: We are expanding the matter and gravitational field around a static, spherically symmetric configuration and then insert this expansion into the linearized Einstein equations to obtain a system of differential equations that allows solving for the linear perturbations, from which we then extract the tidal deformability.

Applying an external quadrupolar tidal field $\mathcal{E}_{ij}$ to a spherically symmetric star results in it developing a quadrupolar moment $Q_{ij}$ as a response. At linear order, this response is proportional to the applied tidal field, such that $Q_{ij} = -\lambda_{\textrm{tidal}} \mathcal{E}_{ij}$, where $\lambda_{\textrm{tidal}}$ is the tidal deformability. The induced quadrupolar moment modifies the $g_{tt}$ metric component, such that at leading order in the asymptotic rest frame at large radii \cite{Thorne:1997kt}

\begin{equation}
\label{gtt_expansion_thorne}
    g_{tt} = -1 + \frac{2 M_{\rm tot}}{r} - \mathcal{E}_{ij} x^i x^j \left(1 + \frac{3 \lambda_{\textrm{tidal}}}{r^5} \right),
\end{equation}
where $x^i$ define a Cartesian coordinate system with $r^2 = \delta_{ij} x^i x^j$.

We now turn to explicitly deriving the equations governing the linear perturbations from the linearized Einstein equations. We focus on static, even-parity, and quadrupolar ($l = 2$) metric perturbations, which we denote by $h_{\mu\nu}$. Further, we choose to work in the Regge-Wheeler gauge, in which $h_{\mu\nu}$ takes the form

\begin{align}
    \begin{split}
        &h_{\mu\nu} = Y_{20}(\theta, \varphi) \times \\
        &\text{diag} \left( - e^{v(r)} H_0(r), e^{u(r)} H_2(r), r^2 K(r), r^2 K(r) \sin^2\theta \right), \label{eq:metric_perturbation}
    \end{split}
\raisetag{3\normalbaselineskip} 
\end{align}

\noindent
where $H_0$, $H_2$ and $K$ describe the radial dependence of each perturbed metric component and $Y_{20}$ is the $(l, m) = (2, 0)$ spherical harmonic. At the same time, we expand the scalar field. We denote the first-order perturbation as $\delta \Phi$ and use the same ansatz as \cite{Sennett:2017etc}, such that

\begin{align}
    \begin{split}
        \delta \Phi(t, r, \theta, \varphi) = \phi_1(r) \frac{e^{- i \omega t}}{r} Y_{20}(\theta, \varphi),
    \end{split}
\end{align}

\noindent
where the same time dependence was chosen for the perturbations in order to ensure that the energy-momentum tensor remains static. We can obtain a set of differential equations that relate the perturbations to the background solutions by expanding the Einstein equations to first order in $h_{\mu\nu}$ and $\phi_1$.

\begin{widetext}
Inserting this expansion into the Klein-Gordon equation (\eqref{KleinGordonEquation}) and only keeping terms linear in the perturbations results in

\begin{align}
\label{phiperturbation}
    \begin{split}
        \phi_1'' &= \frac{u' - v'}{2} \phi_1' + \left[ -2 \phi_0' - r \phi_0'' + \frac{v' + u'}{2} r \phi_0' + \omega^2 r \phi_0 e^{u - v} \right] H_0 \\ 
        &+ \left[ \frac{6 e^u}{r^2} + \frac{v' - u'}{2 r} + 16 \pi \phi_0'^2 + e^u \bigg(V'(\phi_0^2) + 2 \phi_0^2 V''(\phi_0^2) - \omega^2 e^{- v} \right) \bigg] \phi_1 .
    \end{split}
\end{align}

\noindent
Similarly, we expand the Einstein equations, i.e.~we look at $\delta G_{\mu\nu} = 8 \pi \delta T_{\mu\nu}$. The perturbed energy-momentum tensor of the fermionic part is written as $\delta T_{\nu}^{\mu\text{(NS)}} = \text{diag}(- \delta P / c_s^2, \delta P, \delta P, \delta P)$, where we used $\delta e = \delta P\, \partial e / \partial P = \delta P / c_s^2$, with $c_s$ the sound speed. The perturbed energy-momentum tensor of the scalar field is computed by expanding \eqref{scalarfield::EMT}. Subtracting the $\theta \theta$ from the $\phi \phi$ component of the perturbed Einstein equations reveals $H_2(r) = - H_0(r)$. Adding the $\theta \theta$ component to the $\phi \phi$ component allows to obtain an expression for $\delta P$, which can be substituted into the $tt$ minus the $rr$ component to obtain a differential equation for $H_0$:

\begin{align}
\label{h0perturbation}
    \begin{split}
        &H''_0 + \left[ \frac{v' - u'}{2} + \frac{2}{r} \right] H'_0  \\
        &+ \left[ - 8 \pi \frac{1 + 3\,c_s^2}{c_s^2} \phi_0'^2 + 8 \pi \omega^2 e^{u - v} \frac{c_s^2 - 1}{c_s^2} \phi_0^2 - \frac{u' v'+ v'^2}{2} + v'' + \frac{3 u' + 7 v'}{2r} + \frac{u' + v'}{2r c_s^2 } - \frac{6}{r} e^u \right] H_0 \\
        &= \left[ - \frac{16 \pi}{r} \frac{1 + 3\,c_s^2}{c_s^2} \phi_0'' + \frac{8 \pi}{r} \left(3 u' + v' + \frac{u' - v'}{c_s^2} - \frac{4}{r} \frac{1 + 3\,c_s^2}{c_s^2} \right) \phi_0' + \frac{16 \pi}{r} e^u \left( V'(\phi_0^2) \frac{c_s^2 + 1}{c_s^2} + \omega^2 e^{-v} \frac{c_s^2 - 1}{c_s^2} \right) \phi_0  \right] \phi_1.
    \end{split}
\raisetag{5\normalbaselineskip}
\end{align}

\noindent
Here primes denote derivatives with respect to the coordinate radius $r$. The above equation contains a term depending on $v''$, which is explicitly given by

\begin{align}
    \begin{split}
        v'' ={} & 8 \pi e^u \left( r P u' + r P' + P \right) + 16 \pi r \phi_0' \phi_0'' + 8 \pi \phi_0'^2 + 16 \pi r e^u \left(-V'(\phi_0^2) + \omega^2 e^{-v}\right) \phi_0 \phi_0' \\
        &- 8 \pi e^u V(\phi_0^2) \left( r u' + 1 \right)  + 8 \pi \omega^2 e^u \left[  r (u' - v') e^{-v} + e^{-v} \right] \phi_0^2 + e^u \frac{r u' - 1}{r^2} + \frac{1}{r^2}.
    \end{split}
\end{align}

\end{widetext}

As mentioned in \cite{Sennett:2017etc}, for radii larger than the typical size of the combined system, the differential equation for $H_0$ reduces to

\begin{equation}
    H_0'' + \left( \frac{2}{r} + e^u \frac{2 M}{r^2} \right) H_0' - \left( \frac{6 e^u}{r^2} + e^{2 u} \frac{4 M^2}{r^4} \right) H_0 = 0,
\end{equation}

\noindent
which has a solution in terms of associate Legendre functions

\begin{equation}
    H_0 \approx c_1 Q_2^2 \left( \frac{r}{M} - 1 \right) + c_2 P_2^2 \left( \frac{r}{M} - 1 \right).
\end{equation}

\noindent
Expanding this equation in $r/M$ and matching to \eqref{gtt_expansion_thorne} results in

\begin{align}
    \lambda_\text{tidal} ={}& \frac{16}{15} M^5 (1 - 2 \mathcal{C})^2 [2 + 2 \mathcal{C} (y - 1) - y]&  \label{eq:lambda_tidal}\\
    &\times \{3 (1 - \mathcal{C})^2 [2 - y + 2\mathcal{C} (y - 1)] \log (1 - 2\mathcal{C}) \nonumber \\
    &+ 2\mathcal{C} [6 - 3y + 3\mathcal{C}(5 y - 8)] + 4\mathcal{C}^3 [13 - 11 y \nonumber \\
    &+ \mathcal{C}(3y - 2) + 2 \mathcal{C}^2 (1 + y)] \}^{-1},\nonumber
\end{align}

\noindent
where $y \equiv r_\text{ext} H_0'(r_\text{ext}) / H_0(r_\text{ext})$, $\mathcal{C} \equiv M_\text{ext} / r_\text{ext}$ and $r_\text{ext}$ denotes radial position at which $\lambda_\text{tidal}$ is calculated. The dimensionless tidal deformability is defined as $\Lambda_\text{tidal} := \lambda_\text{tidal} / M_\text{tot}^5$.

In order to determine the behavior of $\phi_1$ and $H_0$ at the origin (and thus determine the initial conditions we have to impose), we expand all quantities around the origin as

\begin{align}
    \begin{split}
        \phi_1(r) = \sum_{i} \phi_{1,i} r^i,
    \end{split}
\end{align}
where $\phi_{1,i}$ are the expansion coefficients that do not have any dependency on the radius. Similarly, we also expand $H_0$ with $H_{0,i}$ as the coefficients. After plugging this expansion into Eqs.~(\ref{phiperturbation}) and (\ref{h0perturbation}) and solving the resulting polynomial equations order by order results in

\begin{align}
    \begin{split}
        \phi_1(r) &= \phi_{1,3} \, r^3 + \mathcal{O} \big( r^5 \big), \\
        H_0(r) &= H_{0,2} \, r^2 + \mathcal{O} \big( r^4 \big). \\
    \end{split}
\end{align}
We additionally impose the boundary conditions

\begin{align}
    \lim_{r\to \infty } \phi_1 = 0.
\end{align}
Now, we can use the fact that \eqref{h0perturbation} is invariant under a simultaneous rescaling of $\phi_1$ and $H_0$. Due to this, we can rescale the equations to automatically have $H_{0,2} = 1$. Similarly to the procedure for $\omega$, we use a bisection algorithm to then find the initial $\phi_{1,3}$ such that the above conditions are fulfilled. $\phi_1$ converges to $0$ just as $\phi_0$, so we also set $\phi_1(r) = 0$ for $r > r_B^*$. This allows us to circumvent the divergence of the perturbations, while having no effect on the tidal deformability, since the equations for $\phi_1$, $H_0$ decouple with $\phi_0 \equiv 0$. Then, the tidal deformability is constant for any $r > r_B^*$ and can easily be extracted.

In case the convergence condition cannot be fulfilled, we follow the procedure  in \cite{Sennett:2017etc} and extract $y$ at $r_\text{ext}$ such that it is a local maximum. Since there are two components in the neutron star at play, there can be multiple local maxima, of which we choose the one at the largest radius. 

The code is publicly available along with examples and the procedures to obtain the results.\footnote{\href{https://github.com/DMGW-Goethe/FBS-Solver}{github.com/DMGW-Goethe/FBS-Solver}}

\section{Results \label{sec:results}}


\begin{figure*}
    \centering
    \includegraphics[width = 0.95\textwidth]{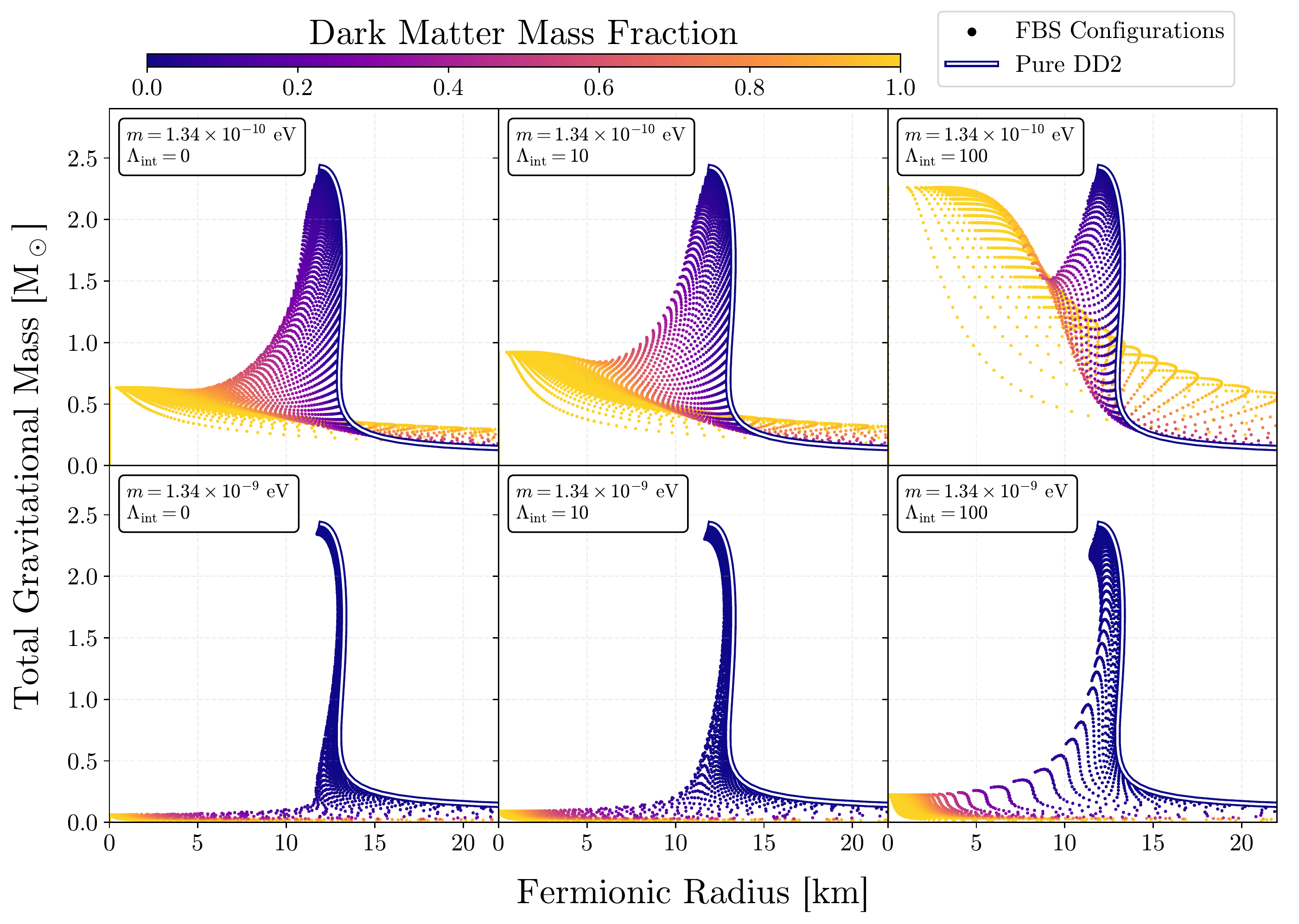}
    \includegraphics[width = 0.95\textwidth]{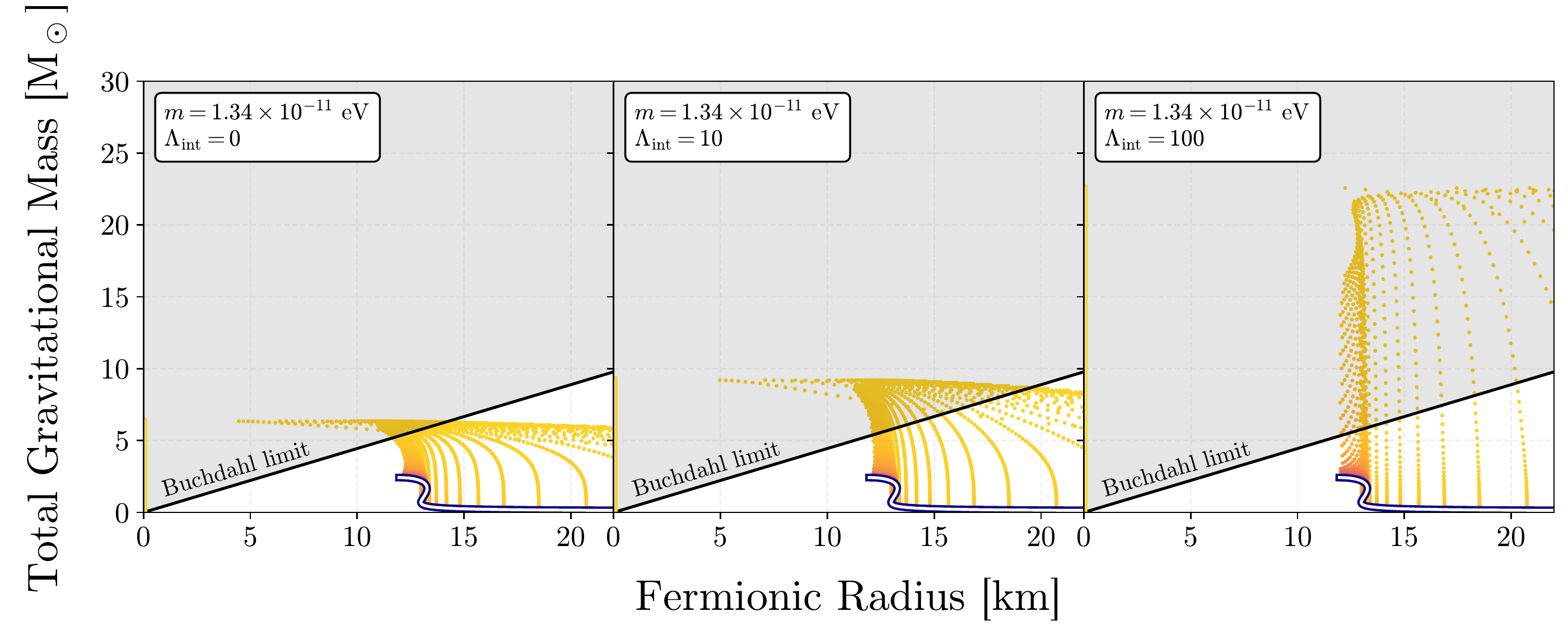}\vskip1cm
    \caption{The relation between total gravitational mass $M$ and the fermionic radius $R_f$ for the FBSs with different DD2 forms mass fractions. The rows correspond to three different bosonic masses $m = \{1, 10, 0.1\}\cdot 1.34 \cross 10^{-10}$\,eV, while the columns correspond to three different $\Lambda_\text{int}= \{0, 10, 100\}$. The EoS we employ for the fermionic part is the \textsc{DD2}. Notice the different scale of the bottom plots. Observing only the fermionic radius of these systems would appear to violate the Buchdahl limit, even though the whole FBS does not. }
    \label{fig:MRgrid}
\end{figure*}

\begin{figure*}
    \centering
    \includegraphics[width = 0.95\textwidth]{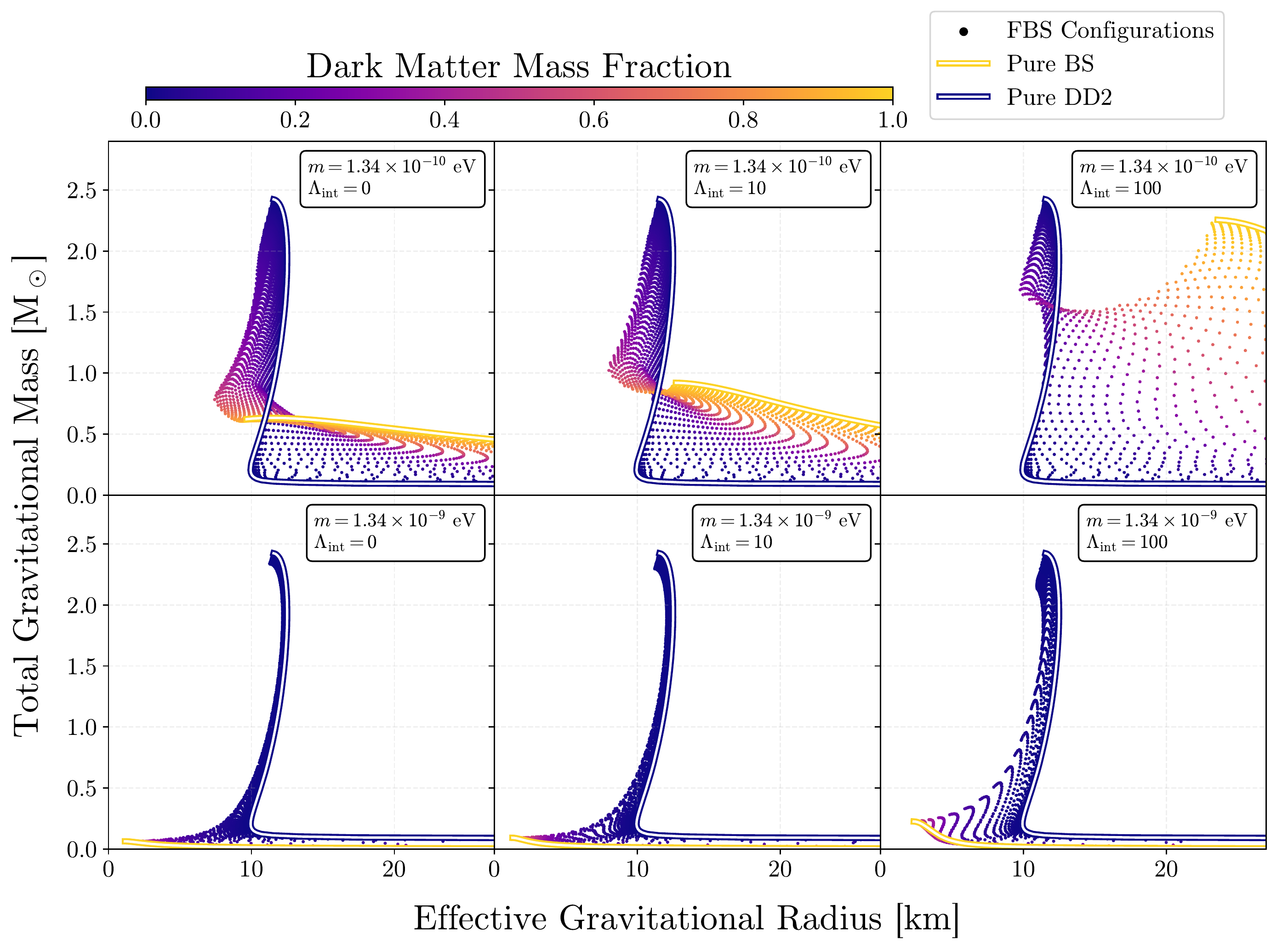}
    \includegraphics[width = 0.95\textwidth]{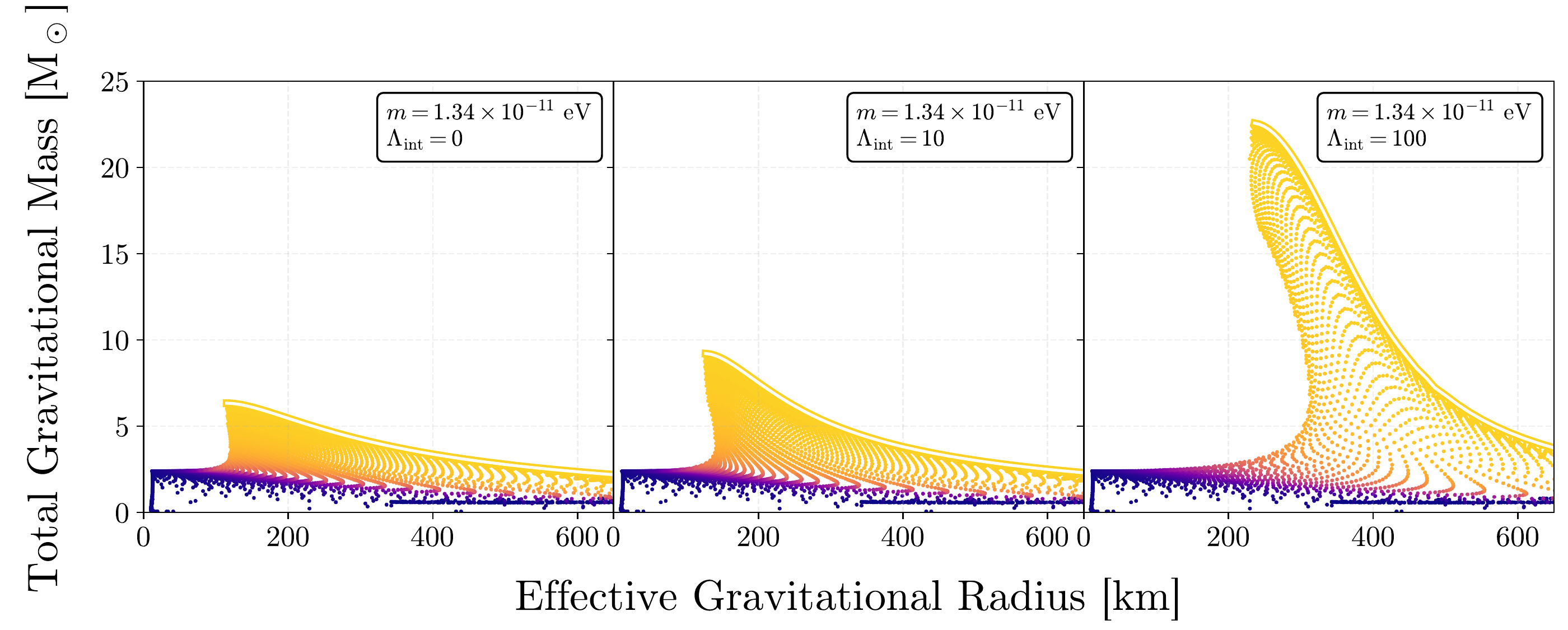}
    \caption{The relation between total gravitational mass $M$ and the effective gravitational radius $R_g$ for the FBSs with different DD2 forms mass fractions. The effective gravitational radius is the radius at which $99\%$ of the rest mass is contained. The rows correspond to three different bosonic masses $m = \{1, 10, 0.1\}\cdot 1.34 \cross 10^{-10}$\,eV, while the columns correspond to three different $\Lambda_\text{int}= \{0, 10, 100\}$. The EoS we employ for the fermionic part is the \textsc{DD2}. In the case of pure neutron stars, the crust has comparatively low density, which makes this effective gravitational radius significantly smaller than the fermionic one. \newline
    Notice the different scales of the bottom plots. For low masses, the bosonic component forms a core and the total compactness of the object increases. For higher masses, the bosonic component forms a cloud and can significantly decrease the compactness of the object.}
    \label{fig:MRggrid}
\end{figure*}

\begin{figure*}
    \centering
    \includegraphics[width = 0.95\textwidth]{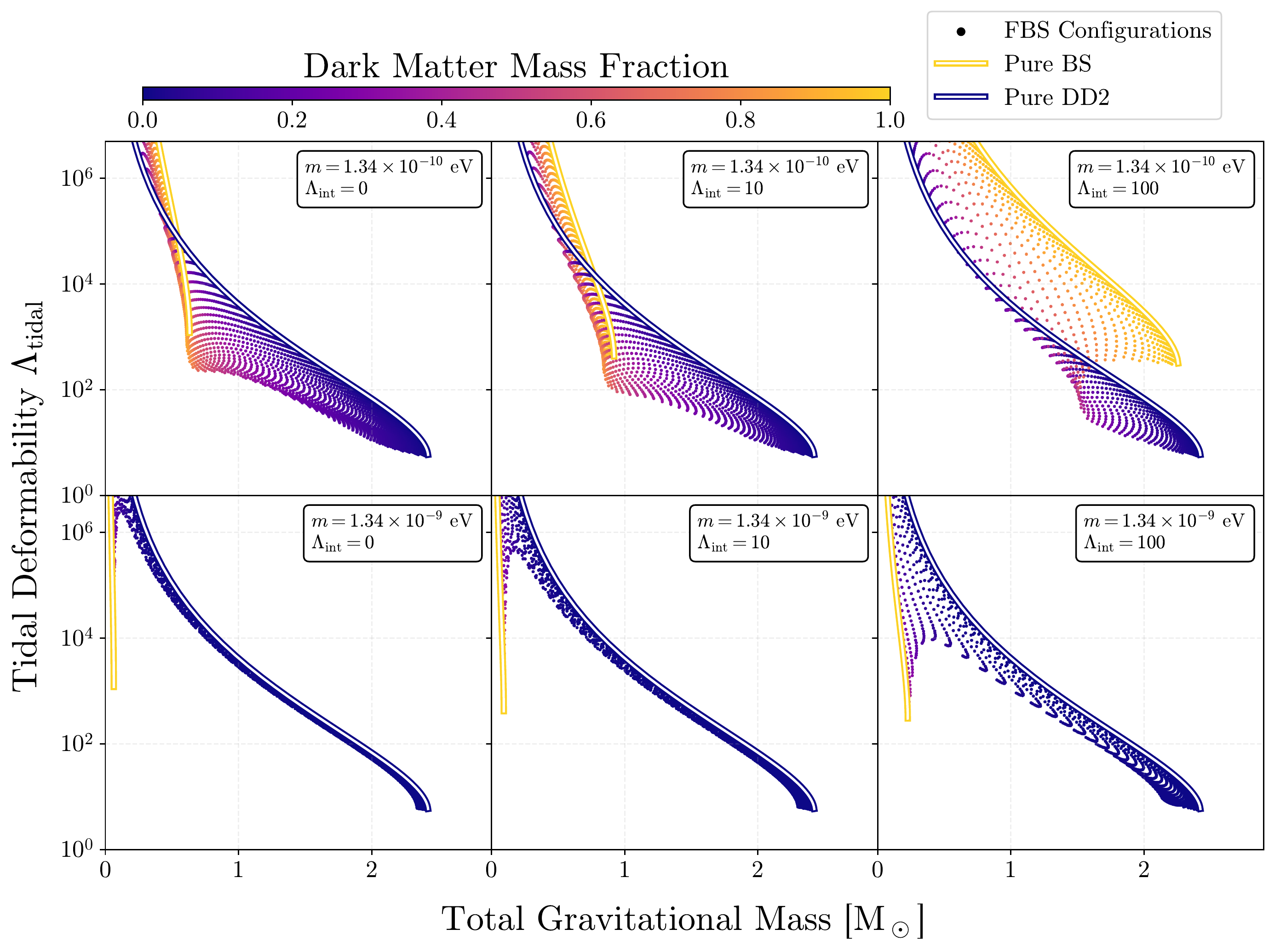}
    \includegraphics[width = 0.95\textwidth]{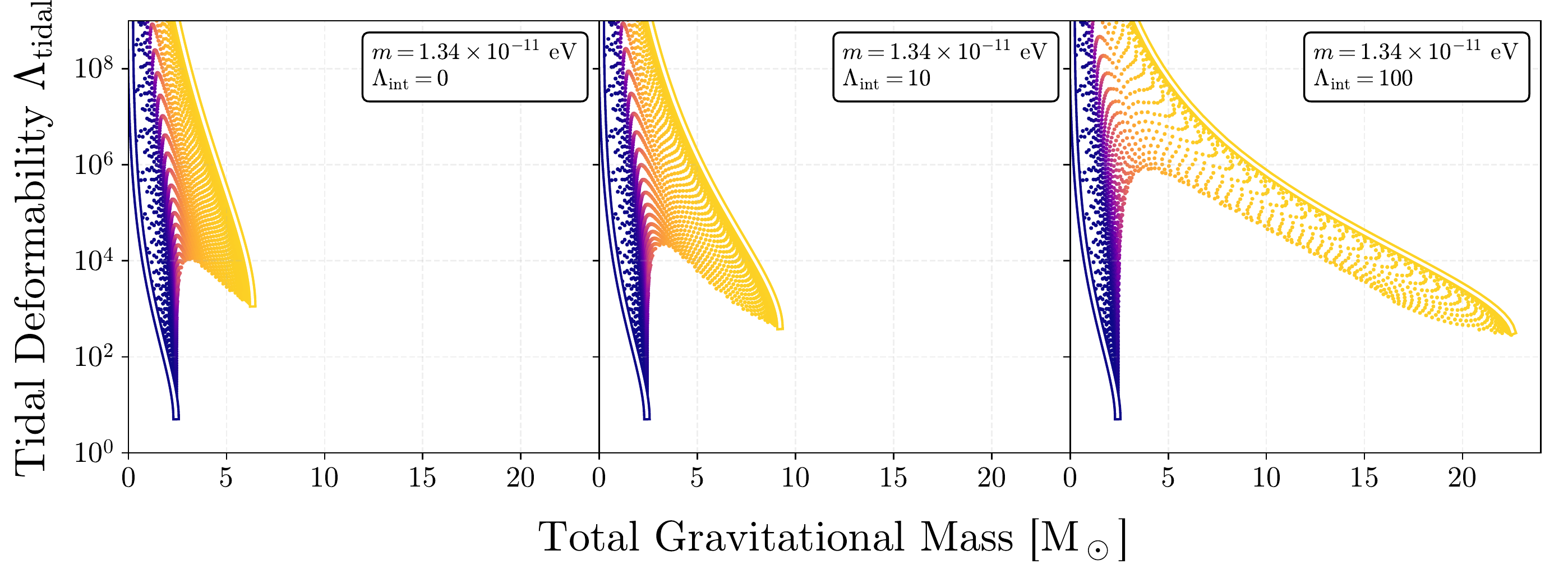}
    \caption{ The relation between dimensionless tidal deformability $\Lambda_\text{tidal}=\lambda_\text{tidal}/M^5$ and total gravitational mass $M$ for the FBSs with different DM mass fraction. The rows correspond to three different bosonic masses $m = \{1, 10, 0.1\}\cdot 1.34 \cross 10^{-10}$\,eV, while the columns correspond to three different interactions strengths $\Lambda_\text{int}= \{0, 10, 100\}$. For the fermionic part, we employ the \textsc{DD2} EoS.}
    \label{fig:MLGrid}
\end{figure*}

We now specialize to a potential that is quartic in the field:

\begin{equation}
    V(\bar{\Phi} \Phi) = m^2 \bar{\Phi} \Phi + \frac{\lambda}{2} (\bar{\Phi} \Phi)^2, \label{eq:potential}
\end{equation}

where $m$ is the particle mass and $\lambda$ is the self-interaction parameter. To allow for easy comparison with previous works, we use the effective interaction parameter $\Lambda_\text{int} = \lambda / (8\pi m^2)$. This was originally introduced in \cite{Colpi:1986ye} to quantify the self-interaction strength, i.e.~for $\Lambda_\text{int} \ll 1$ the total gravitational mass of a pure boson stars scales as $M \propto 1 / m$, while for $\Lambda_\text{int} \gg 1$ we have $M \propto 1 / m^2$. Also, in this regime the stress-energy tensor becomes approximately isotropic, meaning that an EoS might be used to model this case (see sec.~\ref{sec:comparison} below). It is important to keep in mind that $\Lambda_\text{int}$ was introduced in the context of pure boson stars and thus the scaling relations of the total mass are not generally valid for the mixed system, i.e.~FBSs. Nonetheless, we still find it convenient to use it as a general measure to compare different choices of the mass and self-interaction strength.

We investigate nine different models with $m = \{0.1, 1, 10\}\cdot 1.34 \cross 10^{-10}$\,eV and $\Lambda_\text{int} = \{0, 10, 100 \}$. This mass range is chosen such that the Compton wavelength of the bosonic field is half the Schwarzschild radius of the sun, see the explanation in Appendix \ref{app::units}. The range of self-interaction is well within bullet cluster constraints for dark matter, since \cite{Eby:2015hsq, Sagunski:2020spe} 
\begin{align}
    \pi \Lambda_\text{int}^2 m  = \frac{\lambda^2}{64 \pi m^3 } = \frac{\sigma}{m} < {} & 1 \frac{\text{cm}^2}{\text{g}}  \\ \nonumber
    \iff \Lambda_\text{int} < {} & 10^{50} \sqrt{\frac{1.34\cdot 10^{-10}\text{eV}}{m}},
\end{align}
where $\sigma/m$ is the effective cross-section. 

For the fermionic component, we employ the \textsc{DD2} EoS (with electrons) from the CompOSE database \cite{Typel:2009sy, Typel:2013rza}. 
The DD2 EoS is based on a relativistic mean-field model with density dependent coupling constants which has been fitted to the properties of nuclei and results from Brueckner-Hartree-Fock calculations for dense nuclear matter. Thereby, the EoS describes also the EoS of pure neutron matter from chiral effective field theory, see \cite{Kruger:2013kua}. For the purpose of our investigations the particular choice of the nuclear equation of state is not of importance and does not change our conclusions.

\subsection{Mass-Radius Relations and Tidal Deformability}

First, the mass-radius relations are plotted in \figref{fig:MRgrid} and \figref{fig:MRggrid}. 

We use a grid of $\rho_c, \phi_c$ to populate the plots, selecting only the stable configurations as explained in section \ref{sec:equilibrium_solutions}. Each point is colored by the resulting DM mass fraction $N_B/(N_B + N_F)$. Instead of a mass-radius curve, this gives a mass-radius region for the FBSs with different fermionic and bosonic content. Important to note is that in \figref{fig:MRgrid} we plot the fermionic radius, the radius where the fermionic component vanishes. The bosonic radius can be orders of magnitudes larger or smaller, depending on the mass and self-interaction parameter. 
To better understand these objects, we also plot the effective gravitational radius -- the radius at which $99\%$ of the rest mass is contained -- in \figref{fig:MRggrid}. Here, the compactness of the FBS can be inferred. For pure neutron stars with the \textsc{DD2}, the crust has comparatively low density, which makes this effective gravitational radius smaller than the fermionic one.
Which radius is more relevant for a given problem depends on the observation, e.g. the fermionic radius would be crucial for electromagnetic signatures, such as those observed by the NICER telescope. The effective gravitational radius would be more relevant for the inspiral in binary mergers and enters through the compactness and the tidal deformability.

Some general trends can be seen in the figures. Stars dominated by the fermionic part are close to the pure \textsc{DD2} solution, as expected. For stars dominated by the bosonic component, the pure boson star solutions are recovered. 
For $m=\{1,10\}\cdot 1.34 \cross 10^{-10}$\,eV, the regions in \figref{fig:MRgrid} extend to lower masses with similar apparent compactness. These results are consistent with the lines shown in \cite{Di_Giovanni_2021}. A look at \figref{fig:MRggrid} reveals the behavior of these solutions. For $m= 1.34 \cross 10^{-9}$\,eV, the bosonic component is predominantly inside the fermionic one as a DM \textit{core}. For $m=1.34 \cross 10^{-10}$\,eV, the bosonic and fermionic distributions have a similar extent, for low DM mass fraction the compactness is increased, while for higher DM mass fraction the compactness decreases as the DM forms a \textit{cloud}. This is similar to the behavior seen in \cite{Shakeri:2022dwg} for a different mass range, where increasing the DM mass fraction leads to cloud formation.
For $m=1.34 \cross 10^{-11}$\,eV, the bosonic component completely envelops the fermionic one in a cloud and can significantly decrease the compactness of the object (notice the different scales on the x-axis). The apparent compactness of the fermionic part increases on the other hand. Here, only observing the fermionic radius as in \figref{fig:MRgrid} would seem like a violation of GR, as the apparent compactness exceeds the Buchdahl limit of $4/9$.

The relation between tidal deformability and total gravitational mass is plotted in \figref{fig:MLGrid}. Here, we show the dimensionless tidal deformability $\Lambda_\text{tidal} = \lambda_\text{tidal} / M^5$. In blue-bordered lines, the tidal deformability of the \textsc{DD2} EoS is shown, while the tidal deformability of a pure boson star is shown in yellow-bordered lines. The latter agrees with the trend lines shown in \cite{Sennett:2017etc}.

For $m=1.34 \cross 10^{-9}$\,eV, the DM is mostly confined to the inner part of the neutron star as a core and therefore does not affect the tidal deformability significantly. Only for stars completely dominated by DM, the results are close to the pure boson star solutions. For larger interactions $\Lambda_\text{int} \approx 100$, the tidal deformability is decreased.

\begin{figure*}
    \centering
    \includegraphics[width=0.49\textwidth]{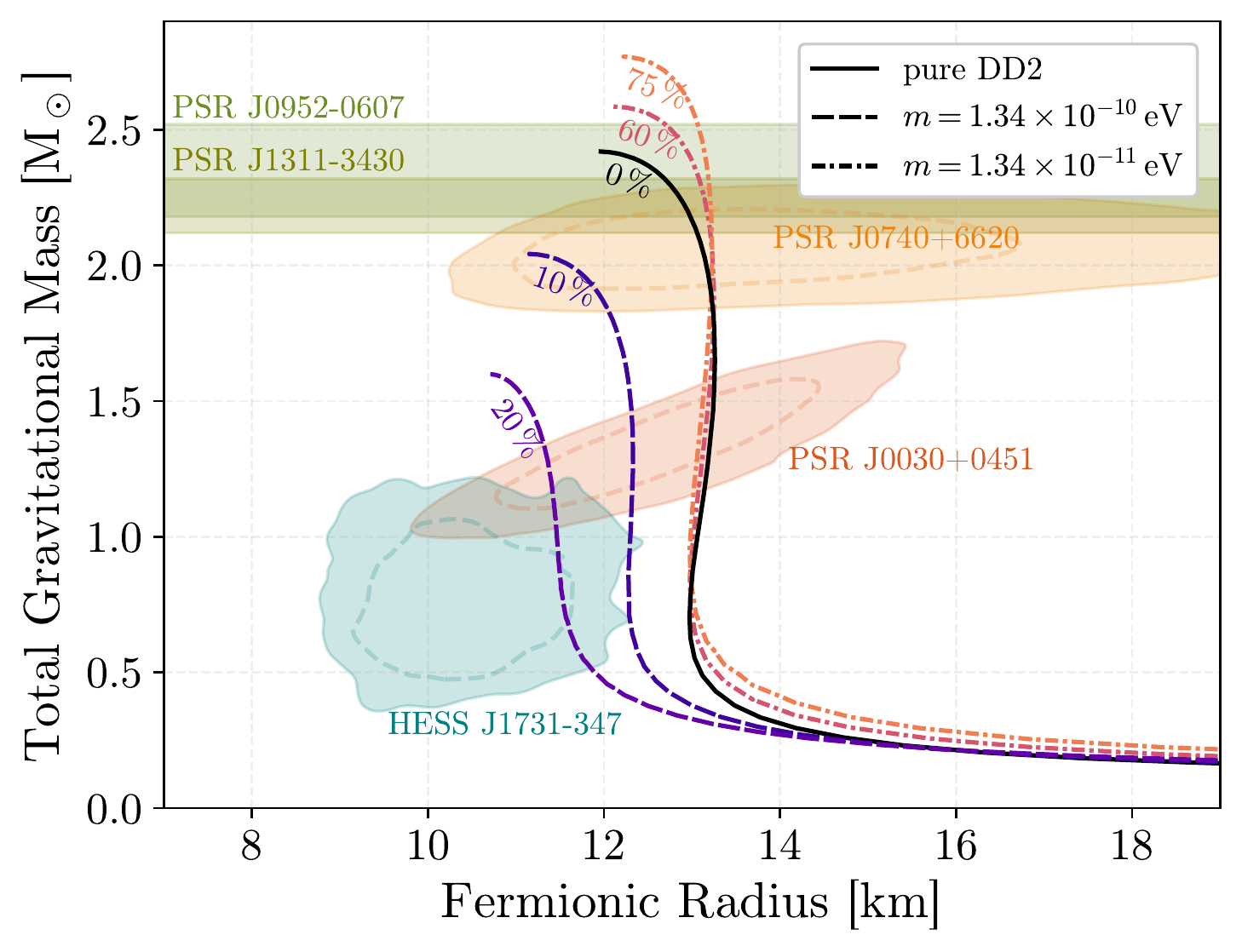}
    \includegraphics[width=0.49\textwidth]{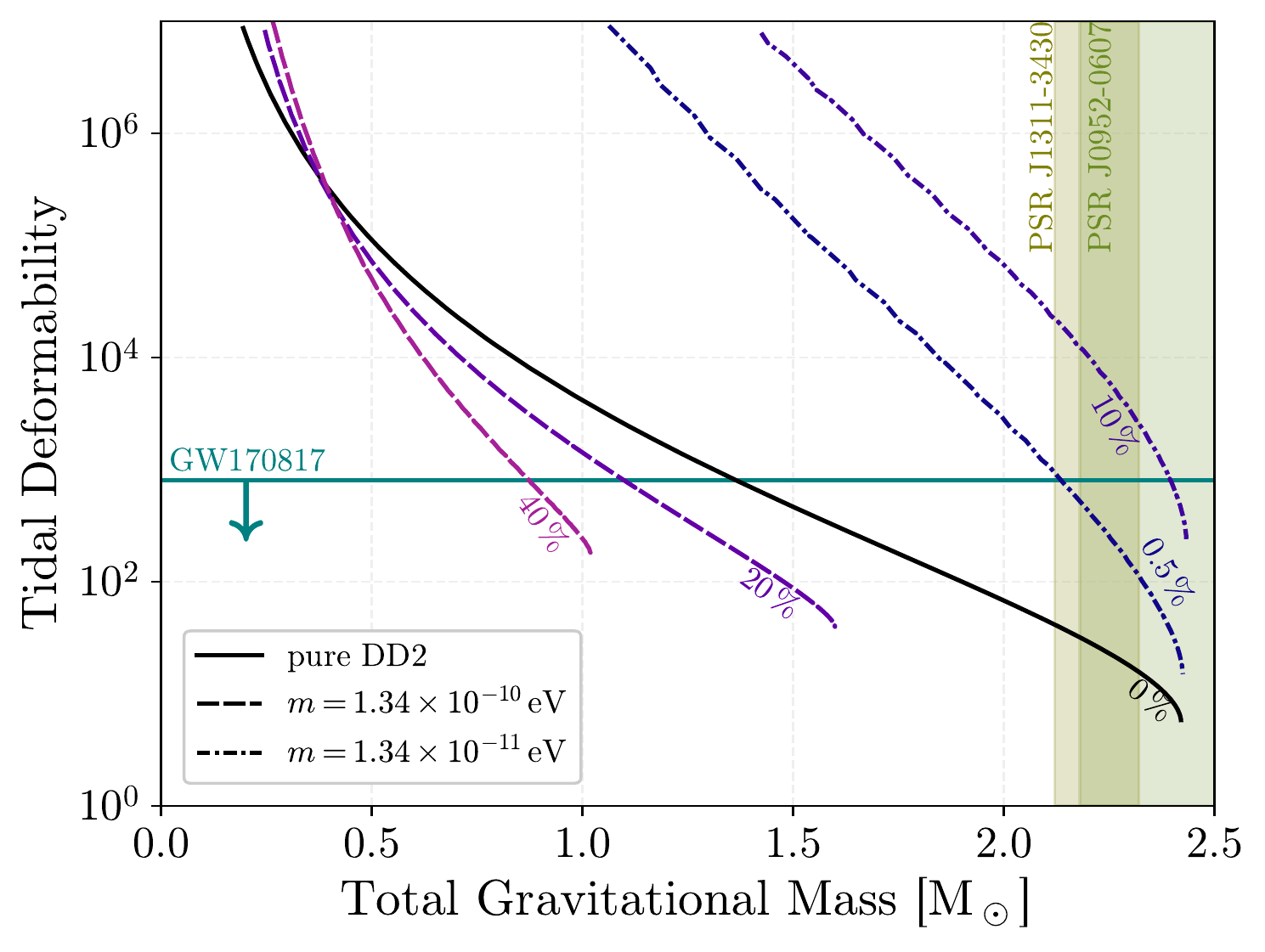}
    
    \caption{\textbf{Left panel:} Resulting mass and radii of FBS for the two cases of $m = 1.34 \times 10^{-10}$\,eV and $m = 1.34 \times 10^{-11}$\,eV shown together with constraints from HESS J1731-347 \cite{Doroshenko2022:aeq}, PSR J0030+0451 \cite{Riley:2019yda}, PSR J0740+6620 \cite{Riley:2021pdl}, PSR J1311-3430 \cite{Kandel:2022qor} and J0952-0607 \cite{Romani:2022jhd}. 
    In both cases, the self-interaction was set to zero and the percentage number denotes the DM mass fraction. For the NICER and HESS measurements, dashed lines display the $1\sigma$, while straight lines show the $2\sigma$ regions. \textbf{Right panel:} Dimensionless tidal deformability for the same set of parameters shown together with the constraint coming from the GW170817 event \cite{Abbott:2018wiz}.}
    \label{fig:measurements}  
\end{figure*}

For $m=1.34 \cross 10^{-11}$\,eV on the other hand, where the bosonic component forms a cloud, there is a significant effect on the tidal deformability. The tidal deformability of boson stars is much higher than the one of purely fermionic ones, so even small amounts of DM can significantly increase the tidal deformability of the FBS. For constant $\rho_c$, the tidal deformability increases orders of magnitude as $\phi_c$ increases. Then, there is a turning point where the tidal deformability decreases while increasing total gravitational mass and converges to the purely bosonic solutions. Overall, this opens up a vast new parameter space, even for small DM mass fractions. While the presence of these bosonic clouds in small quantities would barely be observable in the mass-radius plane, it would clearly affect the tidal deformability even in small quantities, as visible in \figref{fig:measurements}.

For $m=1.34 \cross 10^{-10}$\,eV, the behavior is more dependent on the interaction strength $\Lambda_\text{int}$. For weaker interactions, the tidal deformability stays roughly in the same order of magnitude for constant $\rho_c$, while slowly converging to the pure bosonic solution for increasing $\phi_c$. For stronger interactions, the tidal deformability actually increases as it converges to the bosonic solution, as the bosonic component starts to form a cloud. This behavior is consistent with the observations of \cite{Karkevandi:2021ygv}, where an effective EoS was used for modeling the bosonic component.

\subsection{Comparison to Observational Constraints}

There are measurements of the (fermionic) radius of neutron stars by the NICER telescope, tracking hot spots on their surface with X-ray observations. For the millisecond pulsar PSR J0030+0451 they derive the constraints on the mass $M = 1.34^{+0.15}_{-0.16}$\,M$_\odot$ (68\%) and radius $R= 12.71^{+1.14}_{-1.19}$ (68\%) \cite{Riley:2019yda}. A second, heavier millisecond pulsar PSR J0740+6620 has been measured at $M = 2.07^{+0.07}_{-0.07}$\,M$_\odot$ (68\%) with radius $R= 12.39^{+1.30}_{-0.98}$\,km (68\%) \cite{Riley:2021pdl}.  These measurements constitute only two single points on the mass-radius curve (in the neutron star case) or region (in the FBS case), but it can show which curves/regions would support the existence of such stars.

We plot the posterior distributions of these measurements in \figref{fig:measurements} which should be compared to the regions in \figref{fig:MRgrid}, where the fermionic radius is plotted. 

The FBS solutions with a core become more compact depending on the DM fraction. For higher DM fraction, they are not able to produce the maximum mass required by the PSR measurements. DM cloud solutions on the other hand can easily reach higher maximum masses. This is in accordance with \cite{Shakeri:2022dwg}, who modeled the FBS with an effective EoS and also included the changing photon geodesics due to the DM cloud, and \cite{Rutherford:2022xeb} who performed a Bayesian analysis with the effective EoS. 

Another measurement comes from the supernova remnant HESS J1731-347. Modeling the X-ray spectrum with accurate distance information from GAIA, they report a mass of $M = 0.77^{+0.20}_{-0.17}$\,M$_\odot$ (68\%) with radius $R= 10.4^{+0.86}_{-0.78}$\,km (68\%) \cite{Doroshenko2022:aeq}. This is an unusually light neutron star, which standard star evolution theory struggles to explain, see e.g.~\cite{Stockinger:2020hse}. The authors of \cite{Doroshenko2022:aeq} propose it to be a strange star, but looking at \figref{fig:MRgrid}, this region is also well populated by DM core solutions. Of course, one would have to repeat their analysis with an actual bosonic component to get accurate constraints, which we leave for future work.

Lastly, there is the observation of GW170817, a binary neutron star merger. Reference \cite{Abbott:2018wiz} has derived constraints with minimal assumptions on the nature of the compact objects. They use a mass-weighted linear combination of the individual tidal deformabilities and cite an upper limit of $630$. 
Alternatively, assuming neutron stars with the same EoS, ref.~\cite{Abbott:2018exr} has derived constraints on the tidal deformability with the help of universal relations \cite{Yagi:2013bca, Yagi:2013awa}. These constraints are not perfectly applicable to our case, as the I-Love-Q relations are not necessarily applicable (although they might be \cite{Maselli:2017vfi} -- we leave this for future work) and our two FBS stars might have the same EoS but different DM mass fractions. Nevertheless, we can make some initial guesses. It can be seen that the measurements generally favor lower tidal deformabilities. Extrapolating this to \figref{fig:MLGrid}, this would mean that the DM cloud scenarios with larger tidal deformability are disfavored.  Favored on the other hand are DM core situations, which can lower the tidal deformability. A more thorough analysis might place quantitative constraints on these models, which we leave for future work.

Previous studies using an effective EoS description for the bosonic component reach similar conclusions and have placed initial constraints on different mass ranges, such as \cite{Karkevandi:2021ygv, Giangrandi:2022wht, Sagun:2022ezx}. 

Overall, the different measurements seem complementary, and combining them in a proper analysis might significantly constrain the parameter space. DM cloud solutions can have large tidal deformabilities, even for small DM fractions, these would most likely be observable in LVK measurements. DM core effects on the tidal deformability  on the other hand can hide inside neutron stars and only slightly change the properties even for larger DM fractions. But core solutions can explain the HESS measurement, and assuming different DM fractions for different stars, they are not ruled out by the maximum mass measurement. Of course, these effects are somewhat degenerate with the neutron star EoS. Breaking these degeneracies requires other methods, such as looking at correlations in the galactic DM distribution with the neutron star (FBS) mass distribution \cite{Giangrandi:2022wht, Sagun:2022ezx}.

\section{Comparison with An Effective EoS \label{sec:comparison}}

 
 \begin{figure}[t]
    \includegraphics[width=\textwidth]{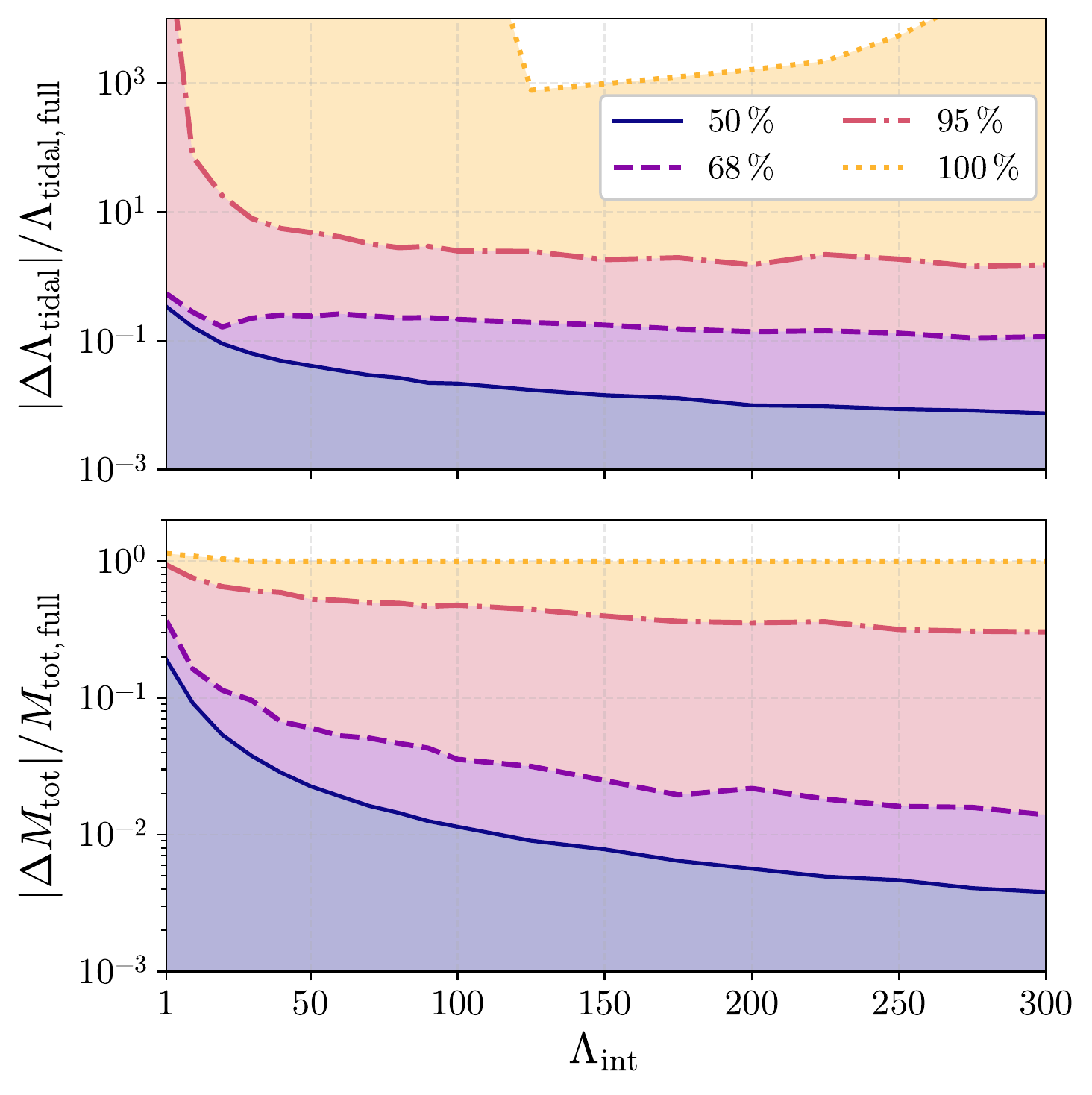}
    \caption{Distribution of the relative error (e.g.~$|\Lambda_{\rm tidal, full} - \Lambda_{\rm tidal, eff}| / \Lambda_{\rm tidal, full}$) of the dimensionless tidal deformability $\Lambda_{\rm tidal}$ (upper panel) and total mass $M_{\mathrm{tot}}$ (lower panel) as a function of $\Lambda_{\rm int}$. For example, the straight line shows the boundary, below which half of the FBS configurations lie, meaning that half of them have a relative error of less than the shown value for a given $\Lambda_{\rm int}$. Subscripts of \textit{full} and \textit{eff} denote quantities obtained from the full system and from the effective EoS, respectively. 
    Only stable FBS were considered for the relative error at a given $\Lambda_{\mathrm{int}}$ and computations were performed for $m=6.7 \times 10^{-11}$\,eV. The agreement between the full system and the effective EoS becomes generally better for large $\Lambda_{\mathrm{int}}$, however, at some point, numerical inaccuracies in the full system dominate the relative error, which starts to be problematic for $\Lambda_{\rm int} \gtrsim 400$. 
    }
    \label{comparison-w-eff-eos:error-comparison-effectiveEOS-fullsys}
\end{figure}

Due to the significant numerical effort associated with solving the full system of equations (eqs.~\ref{eq:klein-gordon-eqation}\,-\,\ref{eq:dPdr-equation}) self-consistently, earlier studies \cite{Colpi:1986ye,Leung:2022wcf} have used an effective EoS $P(e)$ for the scalar field, treating it like a perfect fluid with pressure $P$ and total energy density $e$. The effective EoS was originally derived in \cite{Colpi:1986ye} for the cases where $\Lambda_{\mathrm{int}} = \lambda / 8 \pi m^2 > 0 $ is large (strong self-interactions). It models exclusively the ground state of the scalar field and assumes an isotropic energy-momentum tensor (which is only valid in the given limit). The EoS has the advantage that the scalar field must not be solved for directly, and the evolution equations simplify to the default TOV-equations. The effective EoS is given by

\begin{align}
    P = \frac{4}{9} \rho_0 \left[ \left( 1 + \frac{3}{4} \frac{e}{\rho_0} \right)^{1/2} - 1 \right]^2  , \label{eq:effective-bosonic-eos}
\end{align}
where $\rho_0= m^4/2 \lambda$. Note that our expressions for $\rho_0$ and $\Lambda_{\mathrm{int}}$ deviate from \cite{Colpi:1986ye,Leung:2022wcf} by a factor of two due to the different normalization of the scalar field $\Phi$ and the self-interaction parameter $\lambda$ in the potential (\ref{eq:potential}). The authors of \cite{Leung:2022wcf} used the effective EoS in a two-fluid system of perfect fluids, which interact only gravitationally, to compute the tidal deformability of FBS. In the following, we compare the results obtained from integrating the two-fluid model (see \cite{Leung:2022wcf} for details) and from solving the full system (\ref{eq:klein-gordon-eqation}-\ref{eq:dPdr-equation}). In addition, the tidal deformability is computed as one would for a single-fluid system (details in \cite{Leung:2022wcf}) for the two-fluid model, and as described in section \ref{sec:tidal_deformability} for the full system.

For the initial conditions of the two-fluid model we choose the same conditions as in \cite{Leung:2022wcf}. For better comparability between the full system and the effective EoS, we first want to find an expression relating the scalar field $\phi$ to the energy density $e_{\textrm{eff}}$ of the effective fluid. To derive this relation, we set the $T_{tt}$ component of \eqref{scalarfield::EMT} equal to the $T_{tt}$-component of a perfect fluid (therefore $T^{(\Phi)}_{tt} \stackrel{!}{=} e_{\textrm{eff}} \cdot e^{v}$, and use the approximations used in \cite{Colpi:1986ye} (i.e. neglecting spatial derivatives). We obtain an expression that depends only on the scalar field value $\phi$ (see \eqref{eq:initial-conditions-full-system}), the scalar field mass $m$ and the self-interaction parameter $\lambda$
\begin{equation}
    e_{\textrm{eff}} (\phi) = 2 m^2 \phi^2 + \frac{3}{2} \lambda \phi^4, \label{eq:phi-to-energy-density}
\end{equation}
where $e^{-v} \omega^2  = m^2 + \lambda \phi^2$ was substituted using the Klein-Gordon equation (\ref{eq:klein-gordon-eqation}). Equation (\ref{eq:phi-to-energy-density}) holds for all radii (under the approximations stated above). To get the initial conditions for $e_{\textrm{eff}, c}$, one simply plugs in the corresponding central value of the scalar field $\phi_c$.

\begin{figure*}
    \centering
    \includegraphics[width=0.49\textwidth]{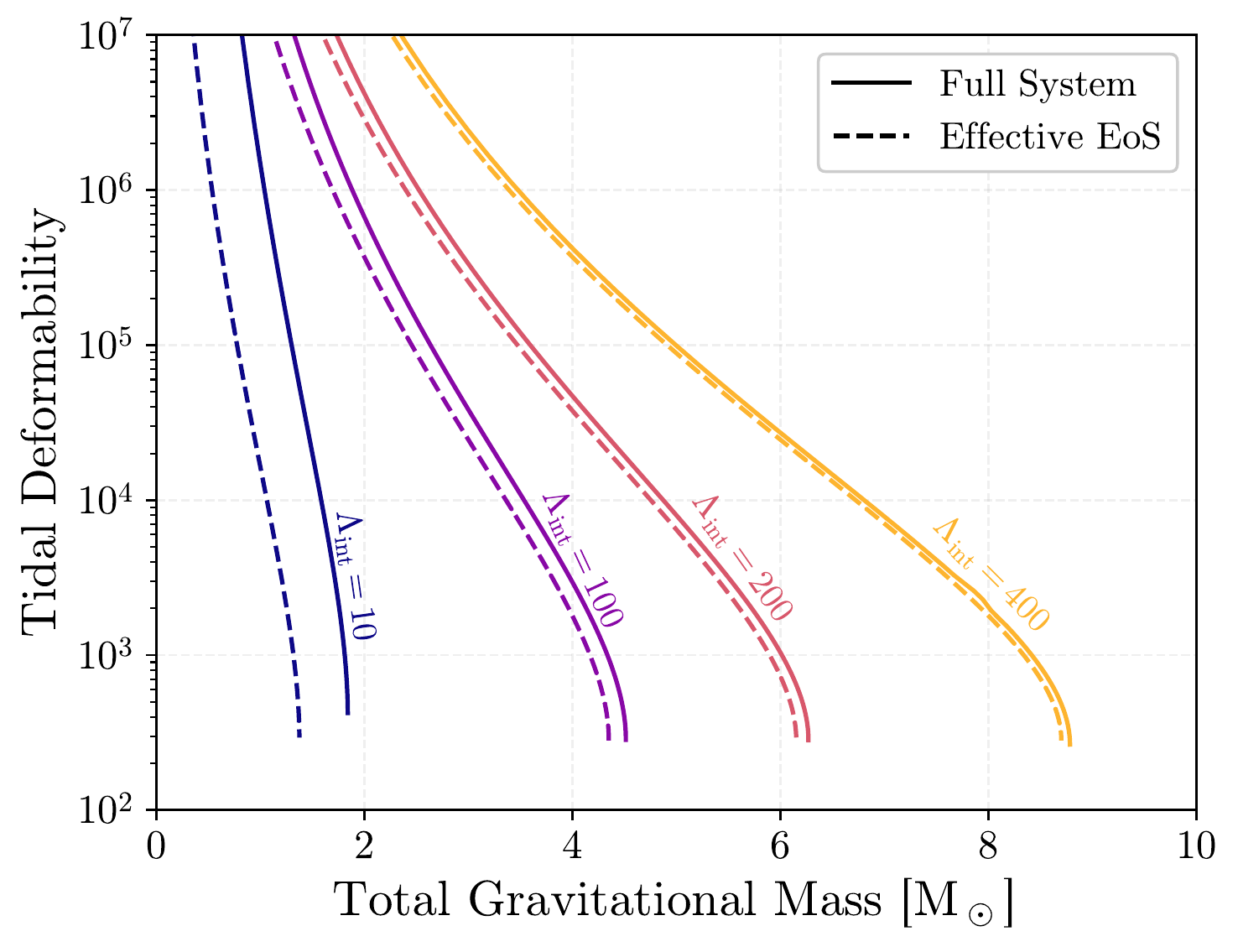}
    \includegraphics[width=0.49\textwidth]{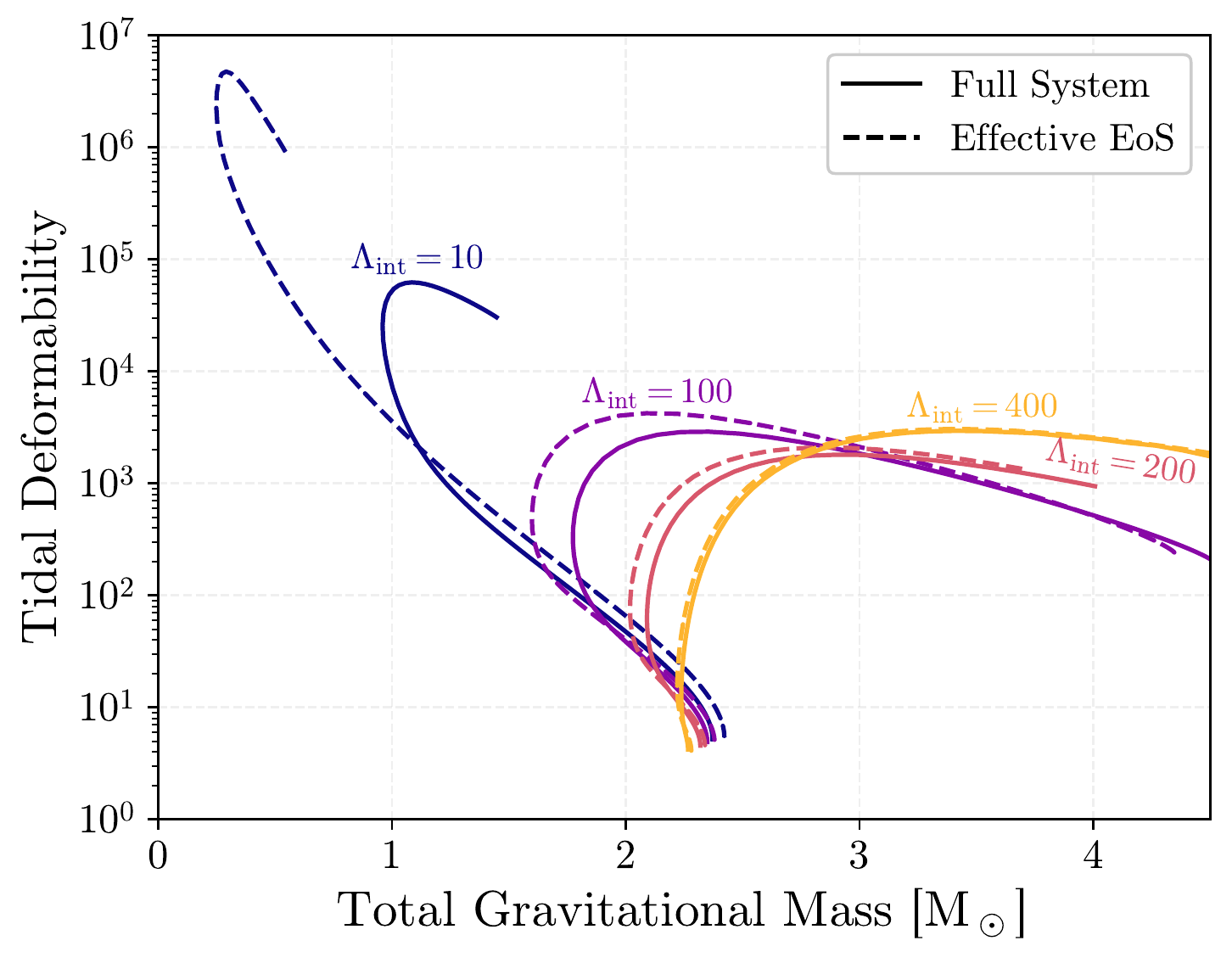}
    \caption{\textbf{Left panel:} Tidal deformability $\Lambda_{\rm tidal}$ plotted against the total gravitational mass $M_{\rm tot}$ for pure BS and various self-interaction strengths $\Lambda_{\mathrm{int}}$. The boson mass is $m=6.7 \times 10^{-11}$\,eV in all cases. The solid lines are the values obtained using the full system eqs.~(\ref{eq:klein-gordon-eqation}-\ref{eq:dPdr-equation}) and the dashed lines are the corresponding solutions using the effective bosonic EoS \eqref{eq:effective-bosonic-eos}. 
    \textbf{Right panel:} Tidal deformability $\Lambda_{\rm tidal}$ with respect to the total gravitational mass $M$ of different FBS for different self-interaction strengths $\Lambda_{\mathrm{int}}$. The boson mass is $m=6.7 \times 10^{-11}$\,eV in all cases. All lines have a constant central value of the scalar field $\phi_c = 0.02$, but different central densities $\rho_c$. Only stars within the stability region are shown. The solid lines are the values obtained using the full system eqs.~(\ref{eq:klein-gordon-eqation}-\ref{eq:dPdr-equation}) and the dashed lines are the corresponding solutions using the effective bosonic EoS \eqref{eq:effective-bosonic-eos}.}
    \label{comparison-w-eff-eos:TLN-full-eff-comparison-Lambda_int}
\end{figure*}

Figure \ref{comparison-w-eff-eos:error-comparison-effectiveEOS-fullsys} shows the relative error $\epsilon_{\mathrm{rel}}$ for the quantities $M_{\mathrm {tot}}$ and the tidal deformability $\Lambda_{\mathrm {tidal}}$, computed using the full system and the effective two-fluid system, with respect to $\Lambda_{\mathrm{int}}$. It can be seen that the errors (the shaded regions) generally decrease for increasing $\Lambda_{\mathrm{int}}$. This is consistent with the assumption that the effective EoS \eqref{eq:effective-bosonic-eos} becomes exact only in the limit of strong self-interactions. For small $\Lambda_{\mathrm{int}}$ the relative error reaches $100\%$ for the total mass and diverges for the tidal deformability. This is to be expected since the total mass converges to zero for pure boson stars when using the effective EoS in the limit $\Lambda_{\mathrm{int}} \rightarrow 0$ (see fig.~2 in \cite{Colpi:1986ye}), while it reaches a constant value when computing the mass using the full system. Likewise, due to the definition of the tidal deformability (see above), a diverging error is to be expected.
For $\Lambda_{\mathrm{int}} \approx 100$ the maximal error of the total mass (tidal deformability) is on the order of $88\,\%$ ($>10^4\,\%$), whereas the lower 95-th percentiles of errors are noticeably smaller at around $< 47\,\%$ ($< 240\,\%$). This means that only $5\%$ of the computed configurations have relative errors higher than $47\,\%$ ($250\,\%$). The median error denoted by the solid blue line is around  $1\,\%$ ($2\,\%$). At $\Lambda_{\mathrm{int}}=300$ the maximal error reaches $85\,\%$ ($>10^4\,\%$) and the median error reaches $0.4\,\%$ ($0.8\,\%$). Asymptotically, the error is constrained by floating-point precision and the inherent error of the effective EoS as compared to the full system.

To gain a better understanding how the effective EoS and the full system compare, we compute the tidal deformability $\Lambda_{\mathrm{tidal}}$ using both systems. The left panel of figure \ref{comparison-w-eff-eos:TLN-full-eff-comparison-Lambda_int} shows the tidal deformability of pure boson stars calculated for different self-interaction strengths $\Lambda_{\mathrm{int}} = \{10, 100, 200, 400 \}$. The solid lines show the solutions using the full system and the dashed lines are the values obtained using the effective EoS. The effective EoS can qualitatively reproduce the solution of the full system, even for small $\Lambda_{\mathrm{int}}$. With increasing lambda, the agreement between full and effective system becomes better. At around $\Lambda_{\mathrm{int}} = 400$, the quantitative agreement reaches a few \% relative difference.

Next, we consider the case for mixed configurations with nonzero scalar field- and central density. The right panel of \figref{comparison-w-eff-eos:TLN-full-eff-comparison-Lambda_int} shows the tidal deformability with respect to the FBS mass. Several curves of constant central scalar field $\phi_c$ were calculated at different $\Lambda_{\mathrm{int}} = \{10, 100, 200, 400 \}$. The choice of constant $\phi_c$ is per se arbitrary but was made for the sake of simpler comparability with future works. The solid lines show the solutions obtained using the full system and the dashed lines were computed with the effective EoS (all other values being equal). With increasing $\Lambda_{\mathrm{int}}$, the solutions using the effective EoS agree with the full system with increasing accuracy. Even though at lower $\Lambda_{\mathrm{int}} < 200$ the deviations are quite large, the qualitative trend is correctly recovered. At $\Lambda_{\mathrm{int}}=400$, both systems produce reasonably similar results (within a few \% of relative difference). This supports the usage of the effective EoS for large $\Lambda_{\mathrm{int}} \gtrsim 400$ also for the computation of the tidal deformability $\Lambda$.

A few notes on the usefulness of the effective EoS \eqref{eq:effective-bosonic-eos} and the two-fluid system: We were able to verify the general notion, that the effective EoS becomes asymptotically more accurate, for most configurations. However, a significant percentage of FBS configurations with high relative errors remain, especially when considering the tidal deformability, where the relative error surpasses 200\% for roughly five percent of all configurations. This is due to the different low mass limits and the definition of the dimensionless tidal deformability. Nevertheless, we conclude that the usage of the effective EoS is justified in the cases where $\Lambda_{\mathrm{int}} \gtrsim 400$, as the errors are acceptable for most (massive) configurations. Of course, solving the full system eqs.~(\ref{eq:klein-gordon-eqation}-\ref{eq:dPdr-equation}) will always yield the exact results in theory. In practice, it can be numerically difficult to integrate the full system at high $\Lambda_{\mathrm{int}} \gtrsim 400$ because (1) the frequency $\omega$ must be tuned up to higher accuracy than what is possible using 64-bit floating-point numbers and (2) increasingly small step-sizes are needed, to solve the equations correctly. During our tests, we could determine that the more relevant constraining factor is the high needed accuracy for $\omega$, rather than the step-size. Smaller initial $\phi_c$ lead to larger bosonic radii $\gg 10$\,km, for which the numerical integration becomes problematic. This concerns $5\,\%$ of the considered configurations. In contrast, the two-fluid system together with the effective bosonic EoS is numerically robust and does not require numerical root-finding for $\omega$, and can manage well with larger numerical step-sizes. With equal step-sizes and initial conditions, the two-fluid system takes around two orders of magnitude less computation time than solving the full system. The speedup can be increased further when considering that the two-fluid system also tolerates larger step-sizes while staying numerically accurate.

\section{Conclusions \label{sec:conclusions}}

In this work, we considered the impact of a complex scalar field on the mass and tidal deformability of neutron stars. The scalar field was assumed to be massive and self-interacting, but to only interact gravitationally with the fermionic neutron star matter. We derived the equations describing the linear perturbations of the combined FBS system induced by the presence of an external gravitational tidal field and numerically solved them to obtain the tidal deformability of the combined system. We found that the scalar field masses $m$ and self-interaction strengths $\lambda$ which result in the \textit{core-like} configurations of the dark matter lead to objects with higher compactness and reduced tidal deformability. This is the case for masses $m \gtrsim 1.34 \times 10^{-10}$\,eV. However, large self-interactions $\lambda$ allow for higher FBS masses or can in some cases result in \textit{cloud-like} configurations. In some of these cases, observing only the fermionic radius would appear to violate the Buchdahl limit.

When comparing the results to available observational data of pulsars, it becomes clear that their uncertainties are currently too large (apart from the pulsar mass measurements) to derive quantitative constraints on the dark matter component. The degeneracy of the effects of DM in the FBS with the EoS poses an additional challenge. As certain DM masses can increase the total mass of the system while leaving the fermionic radius roughly constant, this makes previously excluded EoS possible again, if they appear in a mixed configuration of NS matter and DM. Likewise, the unusually light neutron star HESS J1731-347 is difficult to reconcile with known high-mass pulsar measurements, using a regular EoS. 

The relatively weak constraint from GW170817 on the tidal deformability ($\Lambda_\mathrm{tidal} \leq 800$ at $M_\mathrm{tot} \approx 1.4 \,\mathrm{M}_\odot$) is currently also not strong enough to significantly narrow down the dark matter properties. With the upcoming joint run of LIGO, Virgo and KAGRA, we expect more observational data, which will enable us to derive quantitative constraints. We plan to investigate how to constrain dark matter properties using these observations in the future.

In addition to solving for the scalar field explicitly, we also utilized an effective EoS to describe its contribution to the stress-energy-tensor and reduce the complexity of this model to a two-fluid system. This approach was recently used by \cite{Leung:2022wcf} to compute the tidal deformability. In this work, we compared the result of using the effective EoS to solving the full system of equations. We found that for $m = 6.7 \times 10^{-11}$\,eV and interactions strengths $\Lambda_{\mathrm int} > 300 - 400$ with $\Lambda_\mathrm{int} = \lambda/(8 \pi m^2)$, the usage of the effective EoS is typically justified. We do not expect this conclusion to be dependent on the value of the mass $m$, but rather only on $\Lambda_\mathrm{int}$. Still, even for large values of $\Lambda_\mathrm{int}$, we find a significant number of configurations with relative errors of $> \mathcal{O}(10^2)$.

Finally, it would be interesting to study the exact impact the additional scalar field has on binary merger dynamics. In \cite{Bezares:2019jcb} this was initially studied for a non-self-interacting scalar field. In general, it will be necessary to extend this study to also account for self-interactions, as this can drastically modify the FBS properties and thus impact the observed gravitational wave signal. We will study this in detail in the future.

\begin{acknowledgments}
The authors acknowledge support by the Deutsche Forschungsgemeinschaft (DFG, German Research Foundation) through the CRC-TR 211 `Strong-interaction matter under extreme conditions'– project number 315477589 – TRR 211.
\end{acknowledgments}
\appendix
\section{Units \label{app::units}}

In this work, we considered units in which $c = G = M_\odot = 1$. As a direct consequence, distances are measured in units of $\approx 1.48$\,km, $\hbar \approx 1.2 \times 10^{-76} \neq 1$ and $m_\text{planck} = \sqrt{\hbar c / G} \approx 1.1 \times 10^{-38}$.

We describe the Boson star using the Klein-Gordon equation, which in SI units and flat spacetime reads as $(\square - (mc/\hbar)^2)\phi = 0$. The term $mc/ \hbar$ is the inverse of the reduced Compton-wavelength $\lambda_c = \hbar /mc$, which sets the typical length scale for the system even in the self-gravitating case. Setting it equal to the gravitational radius $GM/c^2$, which in the case of mass-scales of $\sim M_\odot$ is approximately $1.48$\,km, leads to $m = \hbar / c \lambda_c$, which corresponds to $1.34 \times 10^{-10}$\,eV, which then also automatically results in Boson stars with masses $\sim 1$\,M$_\odot$. Previous works such as e.g.~\cite{DiGiovanni:2021ejn} therefore specify the mass of the scalar particle in units of $1.34 \times 10^{-10}$\,eV.

\section{Alternative Conventions}
There is an alternative convention for the metric used in some publications, e.g. \cite{DiGiovanni:2021ejn}, where the spherically symmetric, stationary metric is described by
\begin{equation}
    ds^2 = -\alpha(r)^2 dt^2 + a(r)^2 dr^2 + r^2 (d\theta^2 + \sin^2 \theta d\phi^2)
\end{equation}
The ODEs for the equilibrium solution \eqref{eq:klein-gordon-eqation}-(\ref{eq:dPdr-equation}) are then given by
\begin{align}
    \phi_0'' ={} & \left[ -\frac{\omega^2 a^2}{\alpha^2} + a^2 V' \right]\phi_0 + \left[\frac{a'}{a} - \frac{\alpha'}{\alpha} - \frac{2}{r} \right]\phi_0' \\
    a' ={} & \frac{a}{2} \left[ \frac{1-a^2}{r} + 8\pi r a^2 \left(\frac{\omega^2 \phi^2}{\alpha^2} + V + \frac{{\phi_0'}^2 }{a^2} + \rho(1+\epsilon) \right)  \right] 
\end{align}\\
\begin{align}
    \alpha' ={} & \frac{\alpha}{2} \left[ \frac{a^2 - 1}{r} + 8\pi r a^2 \left(\frac{\omega^2 \phi^2}{\alpha^2} - V + \frac{{\phi_0'}^2 }{a^2} + P \right)  \right] \\
    P' ={}& -\left[ \rho(1+\epsilon) + P \right] \frac{\alpha'}{\alpha}
\end{align}
where $V = V(\phi \bar{\phi})$. Plugging in the quartic potential \eqref{eq:potential}, this would give the same equations as in \cite{DiGiovanni:2021ejn}, except for a different normalization of the field $\phi_0$, which differs by a factor of $\sqrt{2}$.

Making an ansatz for the pertubations as in \eqref{eq:metric_perturbation}
\begin{align}
    \begin{split}
        &h_{\mu\nu} = Y_{20}(\theta, \varphi) \times \\
        &\text{diag} \left( - \alpha(r)^2 H_0(r), a(r)^2 H_2(r), r^2 K(r), r^2 K(r) \sin^2\theta \right), 
    \end{split}
\raisetag{3\normalbaselineskip} 
\end{align}
\begin{widetext}
and performing the same steps leads to the perturbation equations for $H_0, \phi_1$ gives
\begin{align}
    & H_0'' - \left[ \frac{a'}{a} - \frac{\alpha'} {\alpha} - \frac{2}{r}\right]  H_0' \\
    & -  \left[ 8 \pi \omega^2 \phi_0^2 \frac{a^2}{ \alpha^2}  \frac{1- c_s^2}{c_s^2}  + 8\pi {\phi_0'}^2 \frac{1+3c_s^2}{c_s^2} - 2 \frac{\alpha''}{\alpha} + 2 \frac{\alpha' a'}{ \alpha  a} + 4 \frac{\alpha'^2}{\alpha^2} - \frac{a'}{ra} \frac{1+3c_s^2}{c_s^2} - \frac{\alpha'}{r\alpha} \frac{1+7c_s^2}{c_s^2} + 6 \frac{a^2}{r^2} \right]  H_0 \nonumber \\
    &= 16\pi \left[ \omega^2 \phi_0 \frac{a^2}{r\alpha^2} \frac{c_s^2 -1}{c_s^2} +  \phi_0 V' \frac{a^2}{r} \frac{1+c_s^2}{c_s^2} - \frac{\phi_0''}{ r} \frac{1+3c_s^2}{c_s^2} + \phi_0' \frac{a'}{r a} \frac{1+3c_s^2}{c_s^2} +   \phi_0' \frac{\alpha'} {r \alpha} \frac{c_s^2 -1}{c_s^2} - 2  \frac{\phi_0'}{r^2} \frac{1+3c_s^2}{c_s^2} \right]  \phi_1 \nonumber
\end{align}

\begin{align}
    \phi_1'' = \left[ \frac{a'}{a} - \frac{\alpha'}{\alpha} \right]  \phi_1' 
     &+  \left[ - \omega^2 \frac{a^2}{\alpha^2} + 32\pi {\phi_0'}^2 + 2 \phi_0^2 a^2 V'' + a^2 V' - \frac{a'}{ra} + \frac{\alpha'}{r\alpha} + 6\frac{a^2}{r^2} \right] \phi_1 \\
     &+       \left[ \omega^2 r \phi_0 \frac{a^2}{\alpha^2} - r\phi_0'' + \left( r \frac{a'}{a} + r\frac{\alpha'}{\alpha} -2 \right) \phi_0' \right]  H_0 \nonumber                
\end{align}
which need to be complemented with the explicit description for $\alpha''$
\begin{align}
    \alpha'' ={}& 4\pi \omega^2 \left[ 2 r \phi_0^2 a a' + 2r \phi_0 a^2 \phi_0' + \phi_0^2 a^2 \right] \frac{1}{\alpha} 
                 + \left[ 4 \pi r a^2 \left(- \frac{\omega^2 \phi_0^2}{\alpha^2} + P - V + \frac{{\phi_0'}^2}{a^2} \right) 
                                      + \frac{a^2 - 1}{2r} \right]  \alpha' \\ 
                & + \left[ 4 \pi r ( 2 P a a' - 2 V a a'- 2 \phi_0 a^2 \phi_0' V' + a^2 P' + 2 \phi_0' \phi_0'') 
                                   + 4 \pi a^2 (P - V) + 4 \pi {\phi_0'}^2 + \frac{aa'}{r} + \frac{1 - a^2}{2r^2} \right] \alpha \nonumber 
\end{align}
We have implemented both conventions into our code and checked that they give the same results.
\end{widetext}


\bibliographystyle{apsrev4-1}
\bibliography{biblio.bib}{}



\end{document}